\documentclass[pdflatex,sn-mathphys-num]{sn-jnl}

\usepackage[T1]{fontenc}
\usepackage[utf8]{inputenc}
\usepackage{lineno,hyperref}
\modulolinenumbers[20]
\usepackage{graphicx}
\usepackage{color}
\usepackage{tcolorbox}
\usepackage{booktabs}
\usepackage{longtable}
\usepackage{amssymb}
\usepackage{amsmath}
\usepackage{dsfont}
\usepackage{bbm}
\usepackage{lipsum}
\usepackage{enumitem}
\usepackage{multirow}
\usepackage{listings}
\usepackage[table]{xcolor}
\usepackage{float}

\usepackage{listings}
\usepackage{xcolor}

\usepackage{wasysym}
\usepackage{algorithm}
\usepackage{algpseudocode}

\usepackage{subfig}

\lstdefinelanguage{nasm}{
  morekeywords={mov,xor,add,sub,inc,dec,cmp,jmp,jl,jle,jg,jge,je,jne,jz,jnz,
                push,pop,call,ret,nop,lea,test,and,or,shl,shr,imul,idiv,
                short,near,far,byte,word,dword,qword,ptr,ds,cs,ss,es,fs,gs,
                section,global,extern,db,dw,dd,dq},
  morekeywords=[2]{eax,ebx,ecx,edx,esi,edi,ebp,esp,
                   rax,rbx,rcx,rdx,rsi,rdi,rbp,rsp,
                   ax,bx,cx,dx,al,bl,cl,dl,ah,bh,ch,dh},
  sensitive=false,
  morecomment=[l]{;},
  morestring=[b]",
  morestring=[b]',
  alsodigit={0x}
}

\lstdefinestyle{nasmstyle}{
  language=nasm,
  basicstyle=\ttfamily\scriptsize,
  keywordstyle=\color{blue!70!black}\bfseries,
  keywordstyle=[2]\color{purple!80!black},
  commentstyle=\color{gray}\itshape,
  stringstyle=\color{orange!80!black},
  numberstyle=\tiny\color{gray},
  frame=single,
  rulecolor=\color{black!50},
  showstringspaces=false,
  breaklines=true,
  columns=fullflexible,
  keepspaces=true,
  tabsize=2,
  xleftmargin=2pt,
  xrightmargin=2pt
}

\lstdefinestyle{cstyle}{
  language=C,
  basicstyle=\ttfamily\scriptsize,
  keywordstyle=\color{blue!70!black}\bfseries,
  commentstyle=\color{gray}\itshape,
  stringstyle=\color{orange!80!black},
  numberstyle=\tiny\color{gray},
  frame=single,
  rulecolor=\color{black!50},
  showstringspaces=false,
  breaklines=true,
  columns=fullflexible,
  keepspaces=true,
  tabsize=2,
  xleftmargin=2pt,
  xrightmargin=2pt
}

\usepackage{tikz}
\usepackage{adjustbox}
\usepackage{amssymb}
\usetikzlibrary{shapes.geometric, arrows.meta, positioning, fit, backgrounds, shadows, calc, patterns, decorations.pathmorphing}

\usepackage[table,usenames,dvipsnames]{xcolor}
\usepackage[justification=centering]{caption}

\definecolor{techblue}{RGB}{230, 240, 250}       
\definecolor{techborder}{RGB}{70, 130, 180}      
\definecolor{processyellow}{RGB}{255, 242, 204}  
\definecolor{securegreen}{RGB}{200, 230, 200}    
\definecolor{alertred}{RGB}{245, 200, 200}       
\definecolor{darkslate}{RGB}{50, 50, 60}         
\definecolor{softgray}{RGB}{245, 245, 245}       

\usepackage[numbers]{natbib}

\graphicspath{{figs/}}

\begin{document}

\newcommand{\tabalter}[0]{\rowcolors{2}{NavyBlue!12}{NavyBlue!1}}
\newcommand{\tabheader}[0]{\rowcolor{NavyBlue!25}}
\newcommand{\tablight}[0]{\rowcolor{NavyBlue!1}}
\newcommand{\tabdark}[0]{\rowcolor{NavyBlue!12}}

\title[TrapNet]{Applying Graph Analysis for Unsupervised Fast Malware Fingerprinting}
\author*[1]{\fnm{ElMouatez Billah} \sur{Karbab} }\email{mouatez@karbab.net}
\author*[2]{\fnm{Mourad} \sur{Debbabi}}
\affil*[1]{\orgname{Joaan Bin Jassim Academy for Defence Studies}, \orgaddress{\city{Al-Khor}, \country{Qatar}}}
\affil*[2]{\orgname{Concordia University}, \orgaddress{\city{Montreal}, \country{Canada}}}

\abstract{
Malware proliferation is increasing at a tremendous rate, with hundreds of thousands of new samples identified daily. Manual investigation of such a vast amount of malware is an unrealistic, time-consuming, and overwhelming task. To cope with this volume, there is a clear need to develop specialized techniques and efficient tools for preliminary filtering that can group malware based on semantic similarity. In this paper, we propose {\sf TrapNet}, a novel, scalable, and unsupervised framework for malware fingerprinting and grouping. {\sf TrapNet} employs graph community detection techniques for malware fingerprinting and family attribution based on static analysis, as follows: (1) {\sf TrapNet} detects packed binaries and unpacks them using known generic packer tools. (2) From each malware sample, it generates a digest that captures the underlying semantics. Since the digest must be dense, efficient, and suitable for similarity checking, we designed FloatHash (FH), a novel numerical fuzzy hashing technique that produces a short real-valued vector summarizing the underlying assembly items and their order. FH is based on applying Principal Component Analysis (PCA) to ordered assembly items (e.g., opcodes, function calls) extracted from the malware's assembly code. (3) Representing malware with short numerical vectors enables high-performance, large-scale similarity computation, which allows {\sf TrapNet} to build a malware similarity network. (4) Finally, {\sf TrapNet} employs state-of-the-art community detection algorithms to identify dense communities, which represent groups of malware with similar semantics. Our extensive evaluation of {\sf TrapNet} demonstrates its effectiveness in terms of the coverage and purity of the detected communities, while also highlighting its runtime efficiency, which outperforms other state-of-the-art solutions.
}


\maketitle

\section{Introduction}

The number of reported malware samples is growing exponentially, with millions of new samples appearing per month \cite{web_malware_stat_2017}. While 2013 witnessed approximately $30$ million malware samples in the entire year \cite{web_panda_report_2013}, the number for February 2017 alone astonishingly reached $94.6$ million \cite{web_symantec_intel_feb2017}. This phenomenal growth is due to the ease of developing malicious applications, especially by repackaging existing ones to create new variants. While daily malware estimates vary \cite{web_symantec_intel_feb2017, web_panda_report_2013}, they agree on the upward trend. In 2013, new malware was estimated at about $82,000$ per day \cite{web_panda_report_2013}. This figure was estimated at approximately $250,000$ new malware samples per day, a number we use as a reference in our analysis. The major challenge is processing new malware binaries within a limited analysis window. In this context, we address the following questions: (i) How can the triage process for a huge feed of suspicious binaries be made both efficient and effective? (ii) How can this process begin by detecting malicious binaries and (iii) conclude by grouping the samples into families?

\subsection{Problem Statement}

There is an unprecedented need for automated tools, methods, and techniques to tackle the staggering number of daily reported malware samples. For malware triage, unsupervised methods are particularly well-suited as they can cluster malicious applications into groups based on various similarities. This approach allows for more efficient analysis, especially if the resulting groups correspond to well-defined families. The success of such identification depends on the clustering effectiveness. Therefore, we propose a practical and automatic grouping framework that leverages malware similarities without relying on prior knowledge. Common solutions \cite{web_panda_report_2013, web_malware_stat_2017, web_symantec_intel_feb2017} group suspicious binaries based on dynamic analysis of their behavior in a sandbox environment, which provides high resiliency to code mutation techniques. However, this approach has known drawbacks: (i) reduced execution coverage of the analyzed binaries \cite{Hu13MutantX} and (ii) limited scalability due to the need for ample resources. In contrast, we focus on static analysis, as its efficiency makes the triage process scalable enough to handle the tremendous daily growth in malware volume.

\subsection{Solution}

We propose {\sf TrapNet}, a novel framework that employs unsupervised learning for fast malware detection and family clustering based on static features, such as opcodes and function calls. To fingerprint malware, {\sf TrapNet} uses graph community detection to efficiently build a malware similarity network, as shown in Figure \ref{fig_example_network}. It then identifies highly connected nodes (malware), which form suspicious communities. We also propose FloatHash (FH), a novel technique to generate a numerical digest for each malware sample aimed at capturing its underlying logic. The digest has two variants, which are detailed later. FH produces a short vector of real numbers (100 elements in our implementation), which allows for very fast similarity computation. {\sf TrapNet} employs this digest to build the malware similarity network. {\sf TrapNet} combines a set of algorithms and novel techniques to address large-scale malware clustering based on static analysis. Furthermore, it is highly customizable to the needs of security practitioners through a few hyper-parameters.

\begin{scriptsize}
\begin{figure}[H]
    \includegraphics[width=0.75\textwidth, trim=.0cm 0.0cm 0.cm .0cm, clip]{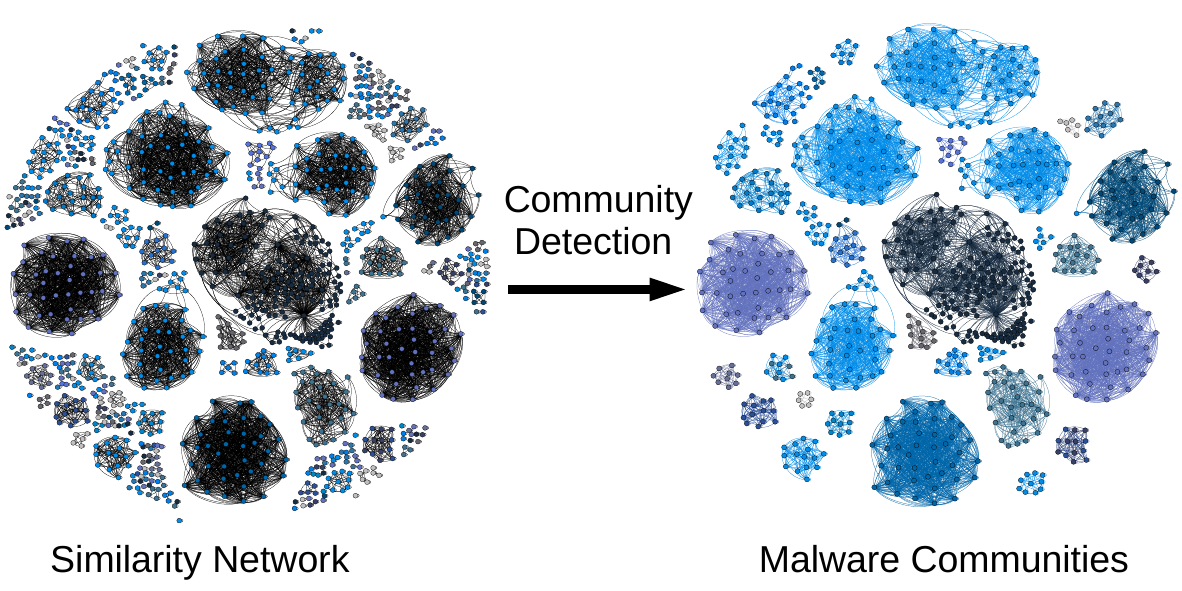}
    \centering
    \caption{From Similarity Network to Malware Communities}
   \label{fig_example_network}
\end{figure}
\end{scriptsize}

\subsection{Contributions}

\begin{enumerate}

	\item We propose {\sf TrapNet}, a novel framework that leverages unsupervised learning for fast malware detection and clustering into families.
{\sf TrapNet} fingerprints malware using a sample similarity network derived via graph community detection.
{\sf TrapNet} is highly customizable to the needs of security analysts (Section \ref{sec_trapnet_framework}).

	\item We propose FloatHash, a novel fuzzy hashing technique that empowers {\sf TrapNet}. FloatHash captures the semantics of any malware sample in a dense vector and is generic and portable enough to be applied to different assembly contents, such as opcodes and function calls (Section \ref{sec_floathash_design}).

	\item We extensively evaluate the effectiveness of {\sf TrapNet} on the Kaggle Microsoft dataset (11.5k samples) and a dataset of recent wild malware (12k samples) (Section \ref{sec_dataset}) for both detection and attribution use cases.
{\sf TrapNet} demonstrates high detection coverage with a high purity percentage in the detected clusters (Section \ref{sec_effectiveness_evaluation}).

	\item {\sf TrapNet} exhibits excellent scalability on a large dataset of $250,000$ recent wild malware samples (Section \ref{sec_dataset}). It required only 12 minutes to cluster the entire dataset, achieving $48\%$ coverage and $82\%$ community purity. {\sf TrapNet} is approximately 15 times faster than MutantX-S \cite{Hu13MutantX} ($130$k samples in 90 minutes), advancing the state-of-the-art in malware clustering from hours to minutes (Section \ref{sec_efficiency_evaluation}).

\end{enumerate}

\section{FloatHash Design}
\label{sec_floathash_design}

\begin{figure*}[H]
  \centering
      \includegraphics[width=0.99\textwidth]{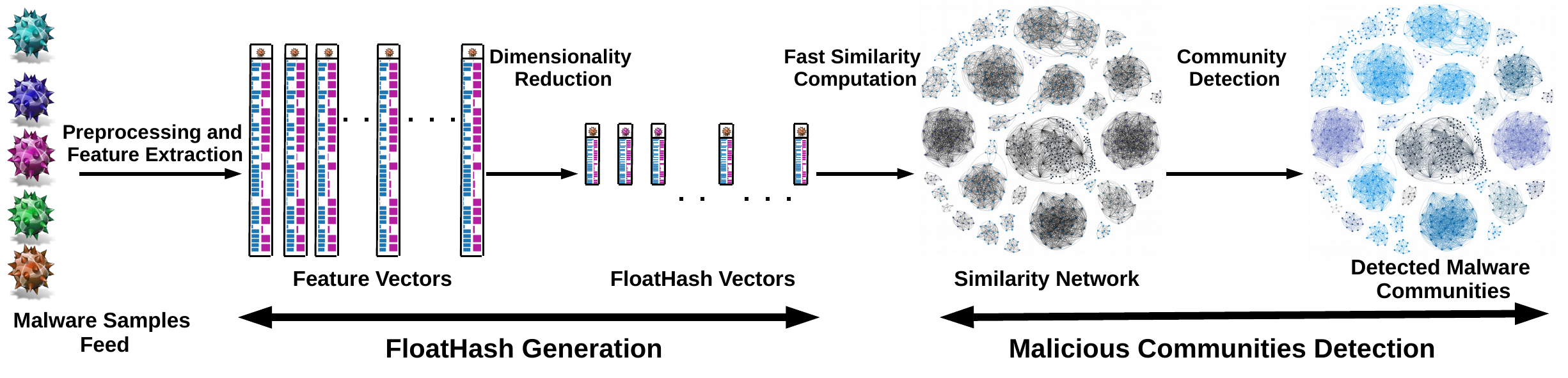}
  \caption{{\sf TrapNet} Approach Overview}
  \label{fig_trapnet_approach_overview}
\end{figure*}

In this section, we present the design of FloatHash (FH), which aims to obtain a one-way compression of a binary's (malware) logic into an embedding vector. For this purpose, FH leverages sequences extracted from the assembly code of the malware sample. In our current implementation, we consider opcode sequences due to their resilience to obfuscation \cite{Hu13MutantX}. We focus on the Win32/x86 assembly instruction set and also consider function invocation sequences. In contrast to BitShred \cite{Jang11BitShred}, where the authors rely on byte sequences from the binary content, we leverage assembly sequences, which are more resilient to common code obfuscation techniques. MutantX-S \cite{Hu13MutantX} leverages opcode {\it N-grams} and feature hashing. In our case, FloatHash considers both opcode and function invocation sequences. The overall FH computation starts by producing a discrete vector from a given sequence (opcodes or functions). Subsequently, we leverage Principal Component Analysis (PCA), a well-known technique for dimensionality reduction, to produce a low-dimensional vector. Note that we generate one embedding vector from the opcode sequence and another from the method invocation sequence for a given malware sample. There are two variants of FH, depending on the generation process of the embedding. In the following subsections, we detail the FH generation process for a given sequence.

\subsection{FH Hashing Variant}
\label{sec_fh_hashing_setup}
In this variant, we parse the opcodes and functions from the assembly code produced by off-the-shelf disassembly tools.

\paragraph{N-grams Computation}
The \textit{N-gram} technique is used to compute contiguous sequences of \textit{N} items from a larger sequence. For our purpose, we use {\it N-grams} to extract sequences employed by the malware sample to distinguish between different samples. Applying {\it N-grams} to opcode and function sequences allows us to capture the underlying malware semantics. We compute a feature vector for each sequence type. A vector $V \in D$ contains a set of {\it N-grams} (along with their number of occurrences) collected from a given sequence, where $D$ represents all possible N-grams, $|D|= \Phi^N$, and $\Phi$ is the language set size.

\paragraph{Feature Hashing} \label{sec_feature_hashing}

The {\it N-gram} technique suffers from its very high dimensionality $D$, which dramatically impacts the computation time and memory required by FH. The complexity of computing the extracted N-gram features increases exponentially with $N$. For example, for \textit{opcode N-grams}, the dimensionality is $D = R^2$ for bi-grams, where $R \approx 400$ is the number of possible opcodes in x86 processors. Similarly, for \textit{3-grams} and \textit{4-grams}, the dimensions are $D=R^3$ and $D=R^4$, respectively.

To address this issue, we leverage the \textit{hashing trick} \cite{qinfeng09hashk} to reduce the high dimensionality of an arbitrary vector to a fixed-size feature vector. More formally, the \textit{hashing trick} reduces a vector $V$ with dimensionality $D=R^N$ to a compressed version with $D=R^M$, where $M \ll N$. Previous work \cite{Weinbergeretal09, qinfeng09hashk} has shown that the hash kernel approximately preserves vector distance. Moreover, the computational cost of using the hashing technique for dimensionality reduction grows linearly with the number of samples and groups, and it helps control the length of the compressed vector in the feature space.

Although feature hashing provides a fixed-length vector, this length is still too high ($2^{16}$ in our case) to leverage very fast approximate similarity computation \cite{NIPS2015_5893}. One might suggest decreasing the size of the feature hashing vector, as it is a tunable parameter. However, this would drastically reduce the quality of the resulting embedding \cite{NIPS2015_5893} due to hash collisions, making it less useful \cite{Hu13MutantX}. For this reason, we leverage PCA to reduce the dimensionality while preserving the maximum amount of information. We aim to reduce the size of $V$ from $2^{16}$ to $S$, where $S \in \{100, 200, \dots\}$ is a hyper-parameter.

\subsection{FH Sequence Variant} 
\label{sec_fh_sequence_setup}

In this variant, we parse the assembly to extract the opcode and function invocation sequences while preserving their order. The process involves two steps:

\paragraph{Code Mapping} We map the sequence of opcodes (and function invocations) that represents the malware sample. More precisely, we replace each opcode with a numerical identifier, resulting in a sequence of numbers. Since assembly instructions do not change frequently, we use all currently known x86 opcodes for Win32/x86 systems to build a mapping dictionary from opcodes to unique numerical identifiers. We then use this dictionary to map the opcode sequence. For function invocation sequences, we developed our dictionary by analyzing $500,000$ Portable Executable (PE) sample files and extracting the most frequent function invocations related to operating system calls. This function dictionary is generated once and then reused for malware fingerprinting. In {\sf TrapNet}, we may encounter unknown function invocations related to third-party libraries. To overcome this, we: (i) used a large dataset ($500,000$ PE samples) to cover most function invocations, and (ii) replace any unknown functions encountered during deployment with a unique identifier.

\paragraph{Sequence Length Unification} The length of the extracted sequences varies between malware samples. Therefore, it is essential to unify the sequence lengths. We use a uniform length hyper-parameter, $L$. There are two cases depending on the original sequence length $l$: (i) If $l > L$, we truncate the sequence, taking only the first $L$ items. (ii) If $l < L$, we pad the sequence with dummy values (zeros). The uniform sequence size $L$ influences the information carried by FH. As a simple rule, the larger $L$ is, the more information the sequence carries. However, a very large $L$ will decrease the efficiency of the overall clustering system. In the current implementation of the FH sequence variant, the discrete vector has a size of $L = 20,000$. At this point, the discrete and uniform sequence is ready for the PCA phase.

\subsection{FH Generation using Dimensionality Reduction} 

Once we have a large discrete vector produced from one of the previous variants, the goal is to lower its dimensionality using techniques such as PCA. We chose to use PCA in the current implementation of FH due to its simplicity. However, FH could leverage other techniques, and we consider the comparison of different information-preserving techniques as future work. We apply PCA to select the top $S=100$ principal components, which constitute the final FloatHash vector. PCA is applied to a matrix where each row represents the discrete vector of a sample. The generated FH for each sample is reusable. This means we can compute a FloatHash model using a matrix from dataset $A$ and use it for similarity computation against a FloatHash vector computed using a matrix from dataset $B$.

\section{TrapNet Framework}
\label{sec_trapnet_framework}

In this section, we present the design of {\sf TrapNet} and describe its main components. {\sf TrapNet} operates in four phases: (1) Binary preparation and preprocessing: preparing the binary for disassembly and preprocessing. (2) FH generation: computing one of the two FH variants, as discussed in Section \ref{sec_floathash_design}. (3) Fast similarity computation: leveraging state-of-the-art approximate nearest neighbor algorithms for FH similarity computation to produce the malware similarity network. (4) Malware community detection: extracting malicious communities by applying state-of-the-art community detection algorithms based on graph modularity.

\subsection{Binary Preparation and Preprocessing}
\label{sec_binary_preparation_preprocessing}

In this phase, {\sf TrapNet} takes a malware dataset (which may include benign samples in a mixed scenario) and produces an opcode and a function invocation sequence for each sample. Initially, {\sf TrapNet} leverages a wide range of known packers to detect and unpack packed samples, relying on tools such as UPX \cite{Wicherski08peHash}. We identify potentially packed samples using tools like PEiD \cite{Gheorghescu05AUTOMATED}, which supports the detection of many packing techniques. Note that the unpacking step is not fully automatic and may require manual intervention in certain situations; we ignore samples with undetected packing, even though such packing could be found through manual analysis. The unpacking step aims to minimize the effect of hidden payloads. Subsequently, {\sf TrapNet} disassembles the samples and preprocesses the assembly code to produce the sequences of opcodes and function invocations.

\subsection{FloatHash Computation}
\label{sec_floathash_computation}

In this phase, we take the sequences from the previous phase, and based on the security practitioner's preference, {\sf TrapNet} computes the chosen FH variant for each malware sample. A malware sample can have multiple FHs depending on the type of sequences used. In the current implementation of {\sf TrapNet}, we consider opcode and function invocation sequences, resulting in two FHs for each malware sample.

\subsection{Similarity Computation}
\label{sec_lsh_similartiy}

Building the similarity network is the backbone of the {\sf TrapNet} framework. We generate this network by computing the pair-wise similarity between the FloatHash vectors of the samples. As a result, we obtain multiple similarity scores corresponding to the number of FH variants. Using various similarity measures provides flexibility and modularity to {\sf TrapNet}. In other words, we can add a new feature vector to the similarity network without disrupting the {\sf TrapNet} process. We can also remove features smoothly, which makes experimenting with feature selection more convenient. More importantly, having multiple similarity measures between static content in the network leads to explainable decisions, allowing a security analyst to track which content makes a pair of samples similar. Since the similarity computation must be highly efficient, we leverage FALCONN \cite{lshforest05bawa}, a tunable high-performance algorithm for similarity computation. FALCONN requires only a few milliseconds per query on a dataset with about 1 million points in around 100 dimensions on a standard desktop machine. We use the well-known Cosine distance:

\begin{equation}
\mbox{cosine-similarity}(v_1, v_2) = \mbox{cos}(\theta) = \frac{v_1 \cdot v_2}{||v_1||||v_2||}
\label{equ:cos}
\end{equation}

\paragraph{Similarity Threshold Heuristic}
\label{similarity_threshold_heuristic}

In the current implementation, for a given pair of malware samples, we compute the similarity between their opcode FH digests and function FH digests separately. We connect their corresponding nodes in the similarity network if they are deemed similar for every FH digest type. This raises the question: How do we define the similarity threshold for a given FH type (opcode and function in our case)? This question becomes more relevant when multiple FH types are used. We could manually search for the optimal threshold, but this would be a difficult task due to the large search space and is not scalable when there are many FH types.

To address this issue, we propose a systematic mechanism to choose the threshold for different FloatHash types, which have different search spaces. The process is as follows: (1) For a given FH type, compute the pair-wise similarity between all FH digests in the dataset. (2) Calculate the average of these similarity values. (3) Use a percentage of the computed average as the threshold for that FH type. This threshold percentage is a fixed hyper-parameter applied to the averages of different FH types. Therefore, tuning the system only requires adjusting this single threshold percentage to meet the security practitioner's requirements regarding the trade-off between purity and coverage.

\subsection{Malware Community Detection}
\label{sec_community_detection}

\begin{figure}[H]
  \centering
      \includegraphics[width=0.6\textwidth]{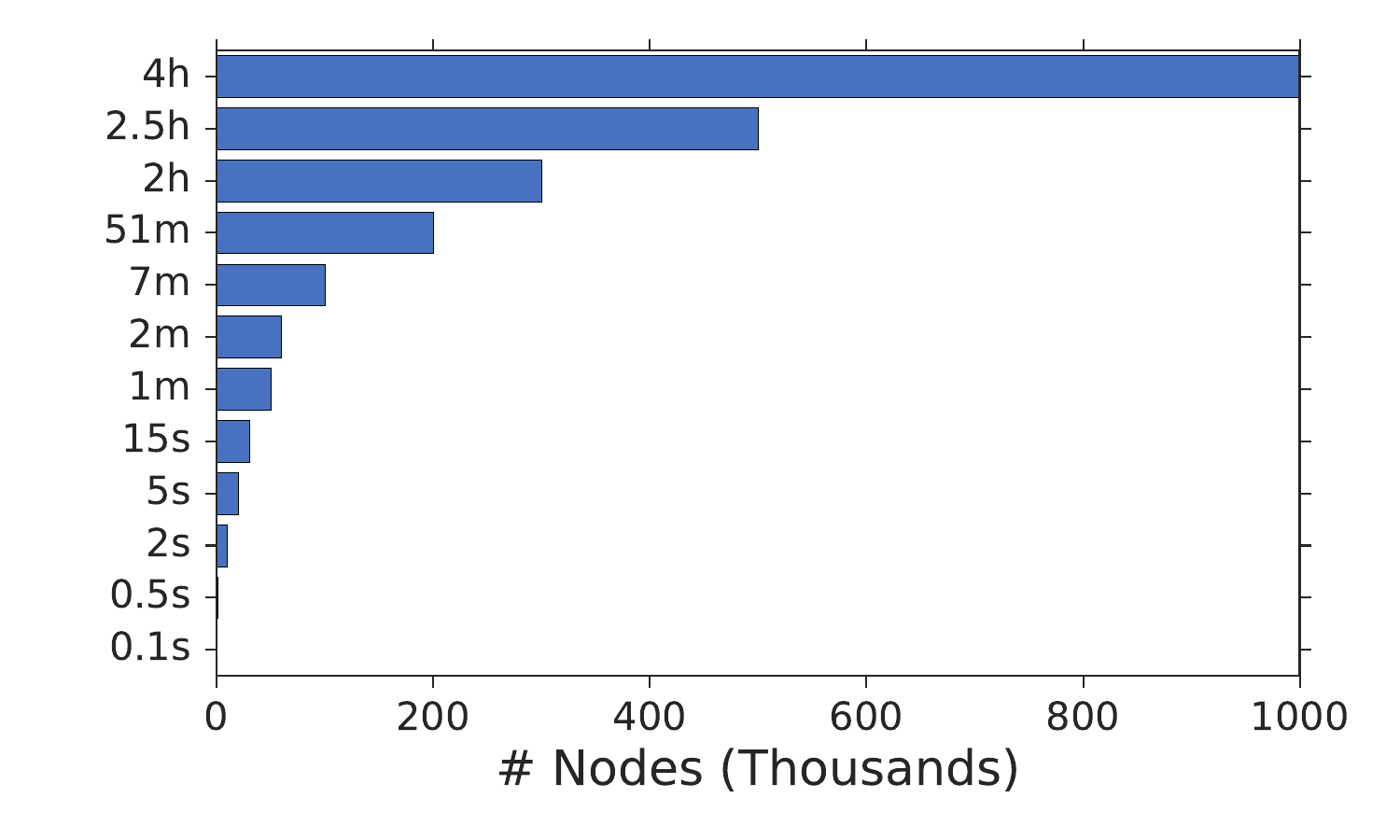}
  \caption{Scalability of The Community Detection Algorithm}
  \label{fig_scale_million}
\end{figure}

A scalable community detection algorithm is essential for extracting communities from our similarity network. For this reason, we endow {\sf TrapNet} with a fast unfolding community detection algorithm \cite{fast08blondel}, which can scale to billions of network links. The algorithm achieves excellent results by optimizing the modularity of communities. Modularity is a scalar value $M \in [-1, 1]$ that measures the density of edges inside a community compared to edges connecting different communities. The algorithm uses an approximation of modularity, since finding the exact optimum is computationally hard \cite{fast08blondel}. Our primary reason for choosing this algorithm is its scalability. As depicted in Figure \ref{fig_scale_million}, we applied it to a one-million-node graph with a relatively higher density than what we encounter in our malware similarity network ($P=0.001$ is the probability of an edge between any two nodes). For completeness, we performed the same experiment on graphs with different densities $P$ (Figure \ref{fig_scale_vs_desity}). As shown in Figure \ref{fig_scale_ultra}, we can detect communities in a $30,000$-node graph with ultra-high density in a relatively short amount of time ($\approx 3$ hours), which is negligible compared to the time required for manual investigation. It is important to note that these experiments test extreme cases compared to real applications, which we present in the evaluation section. For example, on the Microsoft dataset, the community detection algorithm takes less than 10 seconds to compute the similarity network and detect malware communities, as shown in Figure \ref{fig_microsoft_malware_network}.

\begin{scriptsize}
\begin{figure}[H]
    \centering
    \subfloat[$P=0.001$ Medium]{%
        \includegraphics[width=0.4\textwidth]{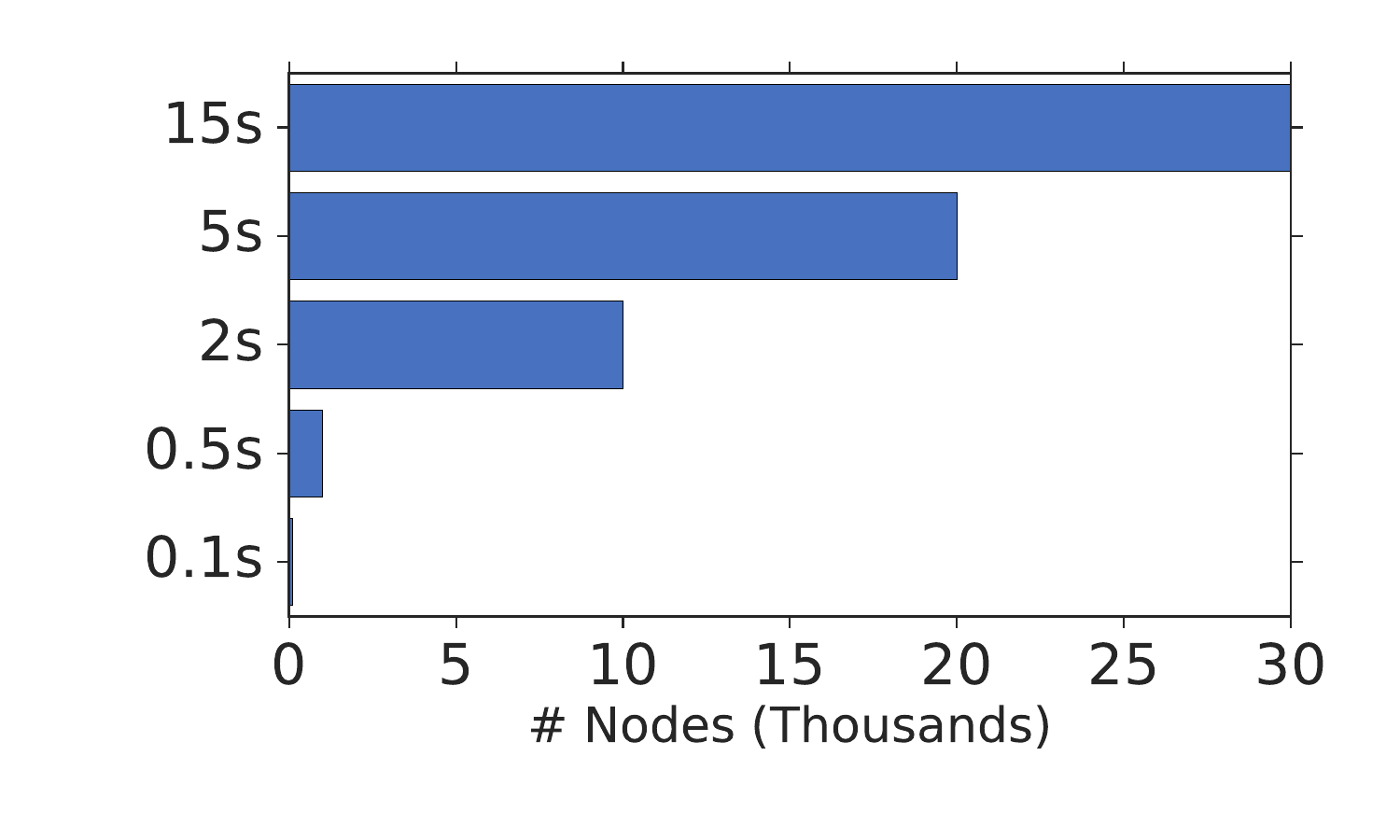}%
        \label{fig_scale_meduim}%
    }
    \hfill
    \subfloat[$P=0.01$ High]{%
        \includegraphics[width=0.4\textwidth]{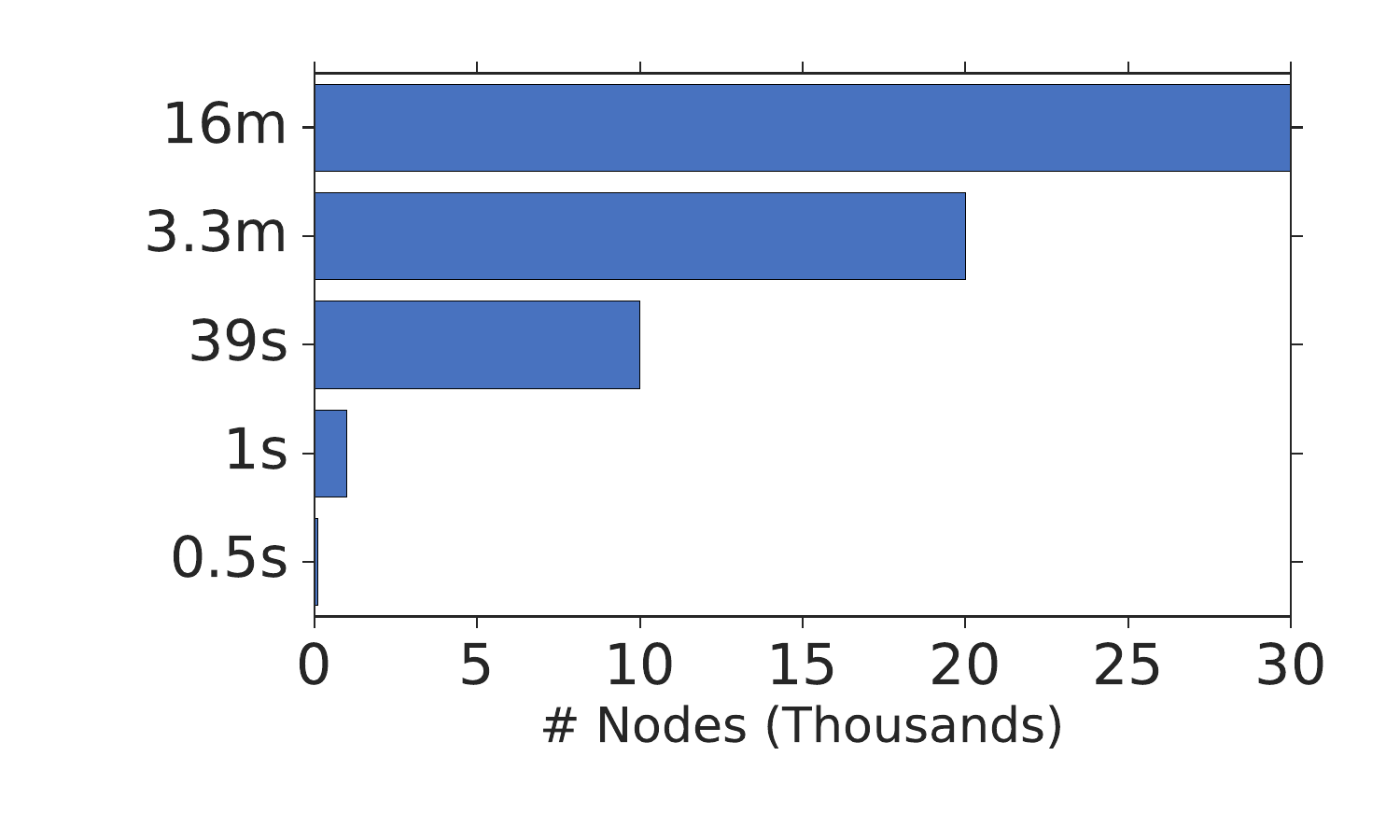}%
        \label{fig_scale_high}%
    }
    \hfill
    \subfloat[$P=0.05$ Very High]{%
        \includegraphics[width=0.4\textwidth]{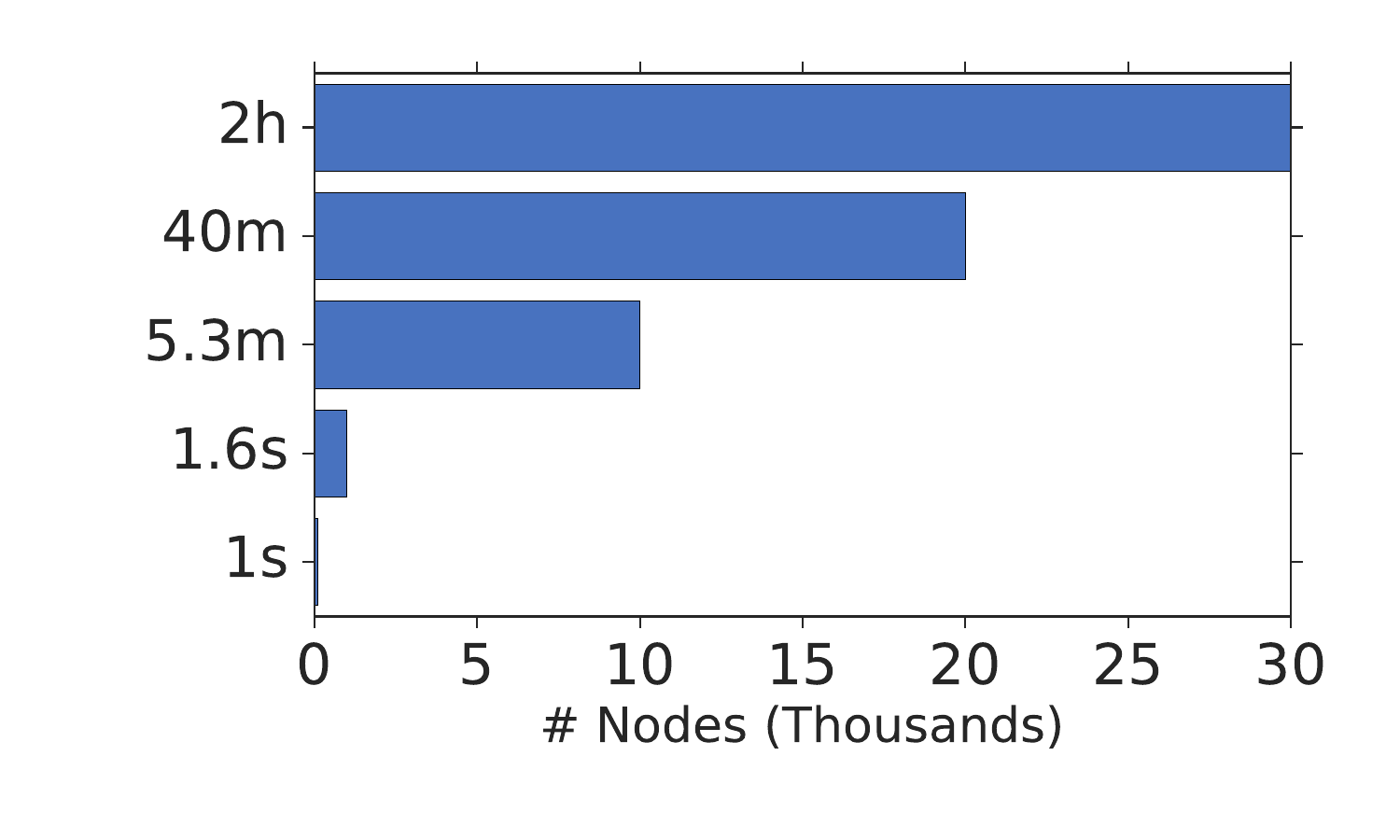}%
        \label{fig_scale_vhigh}%
    }
    \hfill
    \subfloat[$P=0.10$ Ultra High]{%
        \includegraphics[width=0.4\textwidth]{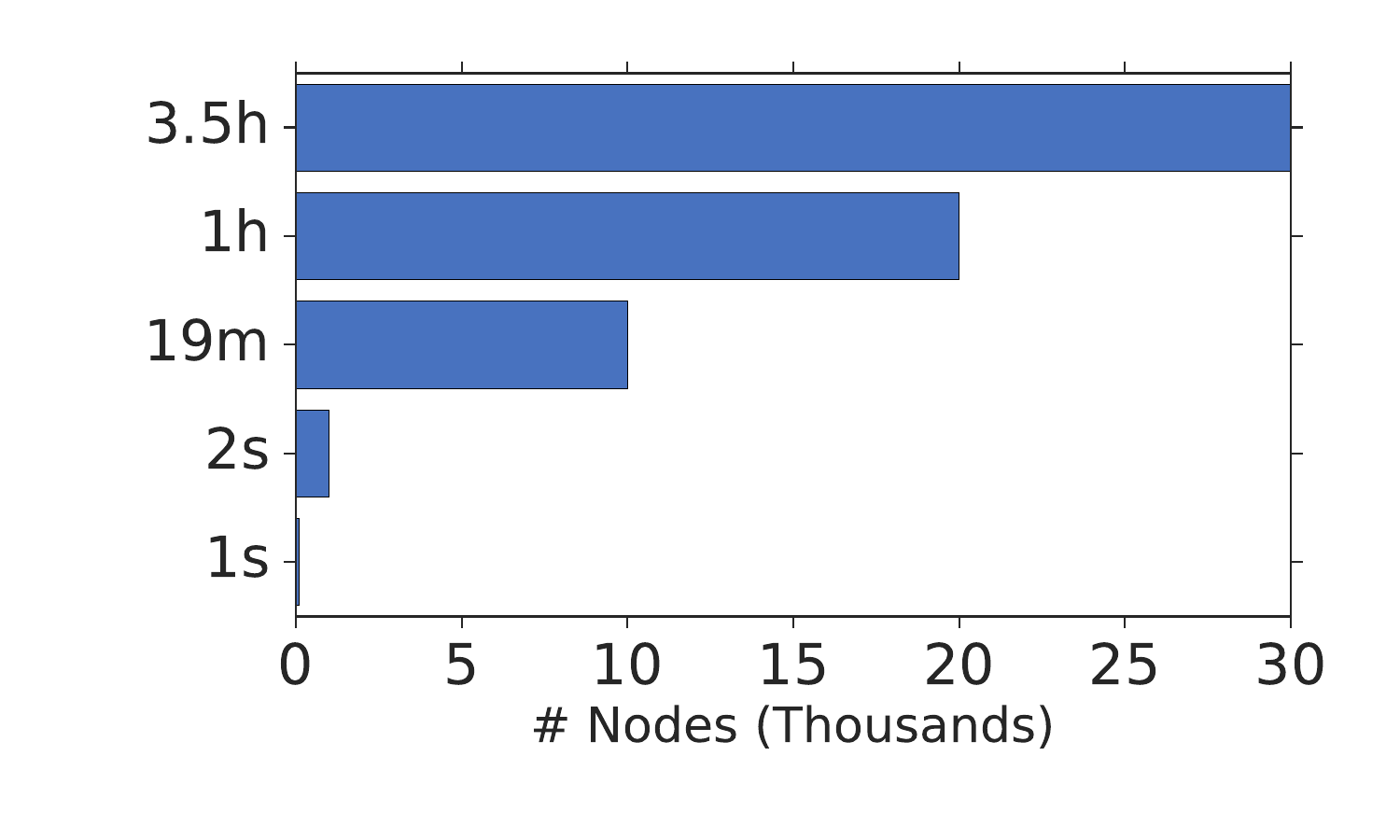}%
        \label{fig_scale_ultra}%
    }
    \caption{Similarity Network Density Versus Scalability}
   \label{fig_scale_vs_desity}
\end{figure}
\end{scriptsize}


\begin{scriptsize}
\begin{figure}[H]
\begin{center}
    \includegraphics[width=0.8\textwidth, trim=0.0cm 0.0cm 0.0cm 0.0cm, clip]{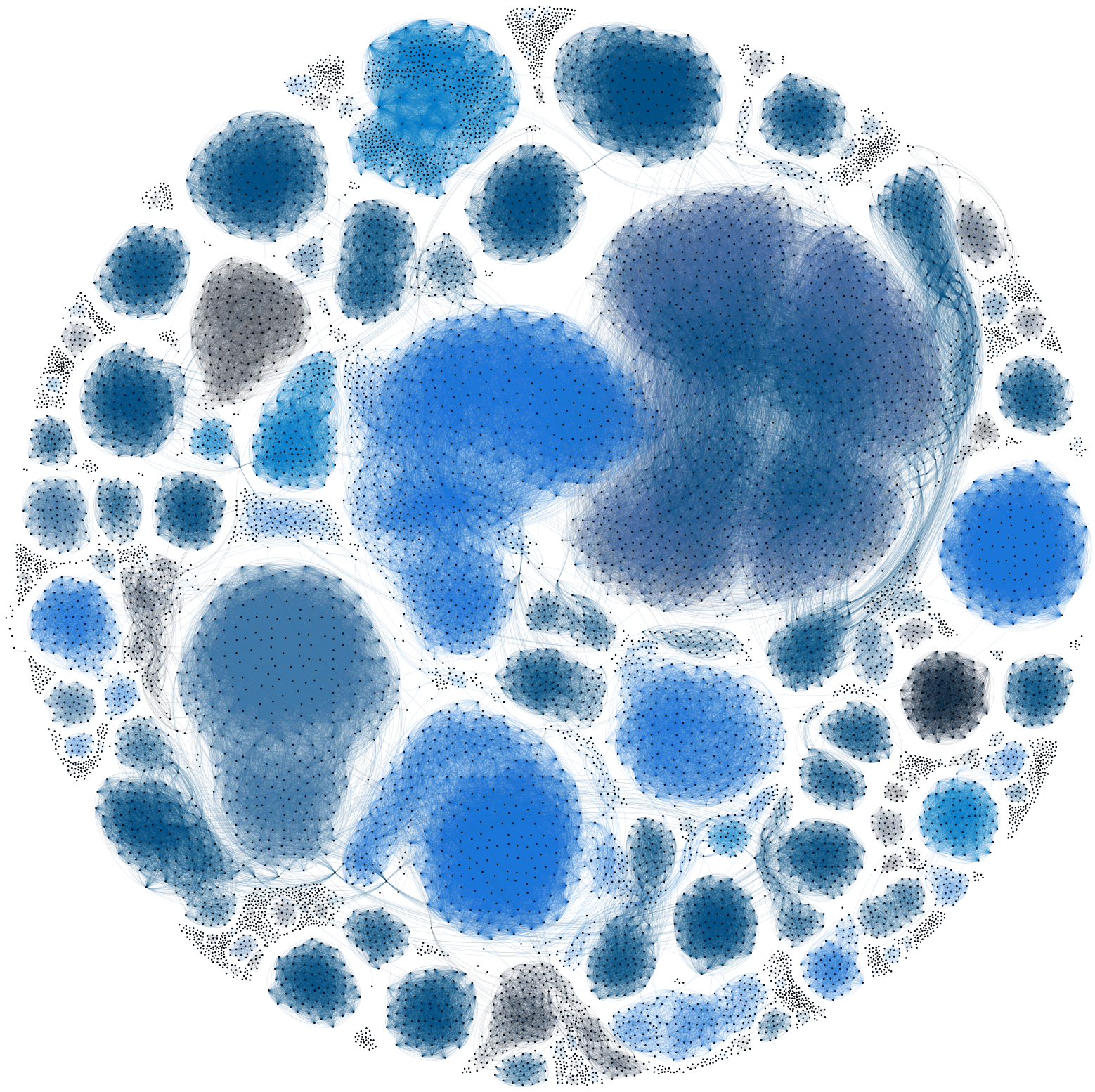}
    \caption{Similarity Network and Detected Communities (Microsoft Dataset)}
   \label{fig_microsoft_malware_network}
\end{center}
\end{figure}
\end{scriptsize}


\section{Implementation}
\label{sec_implementation}

We implemented {\sf TrapNet} using Python and bash scripting. We use IDA Pro\footnote{https://www.hex-rays.com/products/ida/} to disassemble the samples. We then developed custom parsing scripts to extract the opcode and function sequences. For FH generation, which involves computing feature hashing and PCA, we use the Scikit-learn library\footnote{http://scikit-learn.org}. We evaluated {\sf TrapNet} on a server with an Intel(R) Xeon(R) CPU E5-2630 v3 @ 2.40GHz, 128GB of memory, and a 1TB SSD. Although the server has multiple cores, we used only a single core for all {\sf TrapNet} phases except for the similarity computation.

\section{Dataset}
\label{sec_dataset}

We leverage three datasets of different sizes to evaluate {\sf TrapNet}, as shown in Table \ref{tab_evaluation_dataset}. The total number of malware and benign samples is over 263,000.

\begin{table}[!htbp]
\centering
\tabalter
\begin{tabular}{lcc}
\hline\hline
\tabheader
\textbf{Dataset} & \textbf{\# Samples} & \textbf{\# Families} \\
\hline\hline
{\sf TrapNet}  	& 250.0k  & 73 \\
Microsoft       &  10.9k  &  9 \\
Benign         	&   2.2k  &  N/A \\
\hline\hline
\textbf{Total}  &  263.1k &  N/A \\
\hline\hline
\end{tabular}
\caption{Evaluation Datasets}
\label{tab_evaluation_dataset}
\end{table}

First, we use the Kaggle Microsoft dataset to evaluate the effectiveness of {\sf TrapNet}. It contains approximately 11k malware samples from nine families. This dataset comes with the assembly files of the malware samples, so we did not need to perform disassembly. We do not have much information about this dataset besides what is provided, as we do not have the sample hashes to correlate with online sources like VirusTotal\footnote{https://virustotal.com}. To evaluate the mixed scenario, we collected about $2,200$ benign samples, as shown in Table \ref{tab_evaluation_dataset}, representing recent legitimate third-party and system applications from different Windows OS versions.

Second, we built the {\sf TrapNet} dataset, which contains 250k recent malware samples from 73 families (Table \ref{tab_trapnet_malware_dataset}). We use this dataset to evaluate the scalability of {\sf TrapNet}, as the daily number of new malware samples is estimated at 250k \cite{web_malware_variant_2018}. We confirmed the maliciousness of the samples through correlation with VirusTotal, where most samples had more than 35 detections. The family labels were derived from Anti-Virus (AV) scanner reports, which can sometimes be unreliable due to a lack of consensus among AV vendors. For the effectiveness evaluation, we also randomly sampled 12k malware samples from the {\sf TrapNet} dataset while preserving the original number of families.

\begin{table}[!htbp]
\centering

\begin{minipage}{0.48\textwidth}
\centering
\tabalter
\begin{tabular}{clc}
\hline \hline
\tabheader
\#&\textbf{Family} &  \textbf{\#Sample} \\\hline\hline
01&\textbf{allaple  } &57073 \\ \hline
02&\textbf{virut    } &46607 \\ \hline
03&\textbf{nabucur  } &26060 \\ \hline
04&\textbf{mira     } &16383 \\ \hline
05&\textbf{sality   } &13293 \\ \hline
06&\textbf{sivis    } &12643 \\ \hline
07&\textbf{shodi    } &11046 \\ \hline
08&\textbf{ramnit   } & 6055 \\ \hline
09&\textbf{luiha    } & 4556 \\ \hline
10&\textbf{vobfus   } & 4426 \\ \hline
11&\textbf{ipamor   } & 3878 \\ \hline
12&\textbf{loadmoney} & 3637 \\ \hline
13&\textbf{upatre   } & 3023 \\ \hline
14&\textbf{bayrob   } & 2690 \\ \hline
15&\textbf{neshta   } & 2205 \\ \hline
16&\textbf{autoit   } & 1884 \\ \hline
17&\textbf{valla    } & 1855 \\ \hline
18&\textbf{unruy    } & 1847 \\ \hline
19&\textbf{delf     } & 1559 \\ \hline
20&\textbf{zbot     } & 1523 \\ \hline
21&\textbf{madang   } & 1478 \\ \hline
22&\textbf{chir     } & 1417 \\ \hline
23&\textbf{matsnu   } & 1371 \\ \hline
24&\textbf{fynloski } & 1294 \\ \hline
25&\textbf{reveton  } & 1125 \\ \hline
26&\textbf{cutwail  } & 1054 \\ \hline
27&\textbf{expiro   } &  997 \\ \hline
28&\textbf{kryptik  } &  984 \\ \hline
29&\textbf{skyper   } &  952 \\ \hline
30&\textbf{urelas   } &  877 \\ \hline
31&\textbf{fasong   } &  868 \\ \hline
32&\textbf{swrort   } &  818 \\ \hline
33&\textbf{daws     } &  788 \\ \hline
34&\textbf{jadtre   } &  665 \\ \hline
35&\textbf{picsys   } &  641 \\ \hline
36&\textbf{gepys    } &  635 
\\ \hline\hline   
&&\textbf{Total}
\\ \hline\hline
\end{tabular}
\end{minipage}
\begin{minipage}{0.48\textwidth}
\centering
\tabalter
\begin{tabular}{clc}
\hline \hline
\tabheader
\#&\textbf{Family} &  \textbf{\#Sample} \\\hline\hline
37&\textbf{alman     } & 250 \\ \hline
38&\textbf{buzus     } & 248 \\ \hline
39&\textbf{jeefo     } & 244 \\ \hline
40&\textbf{kuluoz    } & 241 \\ \hline
41&\textbf{zaccess   } & 241 \\ \hline
42&\textbf{fesber    } & 229 \\ \hline
43&\textbf{encpk     } & 218 \\ \hline
44&\textbf{pondfull  } & 195 \\ \hline
45&\textbf{slugin    } & 194 \\ \hline
46&\textbf{zegost    } & 187 \\ \hline
47&\textbf{wonton    } & 182 \\ \hline
48&\textbf{waski     } & 163 \\ \hline
49&\textbf{kolabc    } & 150 \\ \hline
50&\textbf{sirefef   } & 145 \\ \hline
51&\textbf{ircbot    } & 139 \\ \hline
52&\textbf{rebhip    } & 137 \\ \hline
53&\textbf{koutodoor } & 119 \\ \hline
54&\textbf{simbot    } & 119 \\ \hline
55&\textbf{waledac   } & 113 \\ \hline
56&\textbf{ardamax   } & 111 \\ \hline
57&\textbf{nimnul    } & 110 \\ \hline
58&\textbf{mabezat   } &  95 \\ \hline
59&\textbf{fakeav    } &  90 \\ \hline
60&\textbf{resdro    } &  81 \\ \hline
61&\textbf{nuqel     } &  81 \\ \hline
62&\textbf{tempedreve} &  72 \\ \hline
63&\textbf{looked    } &  71 \\ \hline
64&\textbf{winwebsec } &  70 \\ \hline
65&\textbf{floxif    } &  70 \\ \hline
66&\textbf{viking    } &  62 \\ \hline
67&\textbf{blakamba  } &  61 \\ \hline
68&\textbf{geral     } &  58 \\ \hline
69&\textbf{tufik     } &  57 \\ \hline
70&\textbf{wauchos   } &  51 \\ \hline
71&\textbf{paskod    } &  44 \\ \hline
72&\textbf{magania   } &  43 
\\ \hline\hline
&\textbf{250000} &
\\ \hline\hline
\end{tabular}
\end{minipage}

\caption{{\sf TrapNet} Malware Dataset}
\label{tab_trapnet_malware_dataset}
\end{table}

\section{Evaluation Metrics}
\label{sec_evaluation_metrics}

In the following sections, we evaluate the effectiveness and efficiency of {\sf TrapNet} on multiple malware datasets. For this purpose, we define the following metrics.


\subsection{Effectiveness Metrics}
\label{sec_effectiveness_metrics}

The following metrics aim to answer these questions: How many malware samples can {\sf TrapNet} group in one iteration? How effective is {\sf TrapNet} at grouping samples of a given family into communities? How many communities does {\sf TrapNet} produce?

\begin{itemize}

\item \textbf{Coverage}: The percentage of samples grouped into communities out of the total dataset size. In a malware-only scenario, higher coverage corresponds to better performance. In a mixed scenario, lower coverage of benign samples is desirable.

\item \textbf{Purity}: The percentage of samples belonging to pure communities (communities with samples from only one family) out of the total number of samples grouped into any community. Higher purity is better in both scenarios. Relying only on coverage and purity is insufficient, as one could achieve perfect scores by placing each sample in its own community. Thus, this metric must be considered alongside the number of detected communities.

\item \textbf{Number of Detected Communities}: The total number of groups generated by {\sf TrapNet}. A smaller number of communities, relative to the number of samples, indicates better clustering. This metric reflects the reduction in manual effort, as an analyst can review communities instead of individual samples.

\item \textbf{Number of Pure Communities}: The number of communities containing samples from a single malware family. This should be examined along with the previous metrics.

\item \textbf{Number of N-Mixed Communities}: The number of communities containing samples from exactly $N$ distinct malware families.

\end{itemize}


\subsection{Efficiency Metrics}
\label{sec_efficiency_metrics}

The following metrics aim to answer these questions: How much time is needed to build the similarity network? How much time is required to extract the communities?

\begin{itemize}

\item \textbf{Similarity Network Building Time}: The time needed to build the malware similarity network via approximate similarity computation on the FH digests.

\item \textbf{Community Detection Time}: The time needed to detect and extract communities from the malware similarity network.

\item \textbf{Total Time}: The sum of the runtimes of the previous two metrics.

\end{itemize}


\section{Effectiveness Evaluation} 
\label{sec_effectiveness_evaluation}

In this section, we assess the effectiveness of {\sf TrapNet} on the Microsoft (MS), {\sf TrapNet} 12k (TN12k), and the complete {\sf TrapNet} (TN250k) datasets. The evaluation covers malware-only and mixed scenarios using both the hashing and sequence variants of FloatHash.

\subsection{Malware-only Scenario}
\label{sec_malware_scenario_eval}

Table \ref{tab_malware_performance_result} summarizes the effectiveness results for the malware-only scenarios. On the Microsoft dataset, {\sf TrapNet} with the FH hashing variant shows excellent results, achieving 82\% coverage (about 8.5k out of 11k samples). Furthermore, 85\% of the grouped malware samples are in pure communities. {\sf TrapNet} produced 155 communities, of which 138 were pure, 11 contained samples from two or three families (possibly due to family mislabeling), and the remaining 6 had more than three families.

\begin{table}[!htbp]
\centering
\tabalter
\begin{tabular}{lcccc}
\hline \hline
\tabheader
 	         & \textbf{Malware} 	          & \textbf{Malware} \\
\tabheader
\textbf{Dataset} & \textbf{Coverage\% / Purity\%} & \textbf{Distribution\footnote{Pure-2Mixed-3Mixed-NMixed} / \#Communities} \\
\hline\hline
\textit{MS (Hashing)}& 82.27\% / 85.0\% &  (138-8-3-6) / 155 \\
\textit{MS (Sequence)}& 44.60\% / 68.0\% &  (96-9-5-5) / 115 \\
\hline
\textit{TN12k (Hashing)}& 45.92\% / 59.0\% & (65-20-12-26) / 123 \\
\textit{TN12k (Sequence)}& 71.15\% / 63.0\% & (91-15-11-9) / 126  \\
\hline
\textit{TN250k (Hashing)}&  0.53\% / 95.0\% & 	(7-1-2-1) / 11 \\
\textit{TN250k (Sequence)}&  48.05\% / 82.0\% &  (946-256-105-97) / 1404 \\
\hline\hline
\end{tabular}
\caption{Malware-only Scenario Performance Results}
\label{tab_malware_performance_result}
\end{table}

The effectiveness of {\sf TrapNet} was also assessed on 12k samples from our {\sf TrapNet} Dataset, which provides a more realistic view of real-world malware because: (1) it contains more families (73), and (2) the samples may be obfuscated, unlike the Microsoft samples. The results on TN12k are shown in Table \ref{tab_malware_performance_result}. Interestingly, {\sf TrapNet} with the sequence variant shows higher coverage and purity compared to the hashing variant.

Surprisingly, when evaluating {\sf TrapNet} on the complete 250k {\sf TrapNet} dataset (Table \ref{tab_malware_performance_result}), we found that the FH sequence variant achieved very good results, with about 50\% coverage and 82\% purity. However, the FH hashing variant performed poorly, grouping only 0.5\% of the dataset, albeit with 95\% purity. We conclude that the FH sequence variant is more scalable and stable than the FH hashing variant. The efficiency evaluation in the next section will further support these findings.


\subsection{Mixed Scenario}
\label{sec_mixed_scenario_eval}

In this section, we consider a dataset of suspicious samples, which is assumed to contain benign files that represent false positives from an earlier triage stage. This assumption is realistic, as the number of false positives can be high in early triage. We focus on the performance of benign sample detection, as the malware detection results are similar to those in the malware-only scenario. For benign samples, lower coverage indicates better performance.

\begin{table}[!htbp]
\tabalter
\begin{tabular}{lcccc}
\hline \hline
\tabheader
 	           & \textbf{Benign} & \textbf{Benign} \\
\tabheader
\textbf{Dataset} & \textbf{Coverage\% / Purity\%} & \textbf{Distribution\footnote{Pure-2Mixed-3Mixed-NMixed} / \#Communities} \\
\hline\hline
\textit{MS (Hashing)}&  20.93\% / 100.0\% & 	(31-0-0-0) / 31 \\
\textit{MS (Sequence)}& 33.77\% / 18.0\% & 	(6-2-4-5) / 17 \\
\hline
\textit{TN12K (Hashing)} & 30.03\% / 65.0\% & 	 (18-3-4-8) / 33 \\
\textit{TN12K (Sequence)} & 65.59\% / 67.0\% & 	 (8-0-1-2) / 11 \\
\hline\hline
\end{tabular}
\caption{Mixed Scenario Performance Results}
\label{tab_mixed_performance_result}
\end{table}


As shown in Table \ref{tab_mixed_performance_result}, {\sf TrapNet} with the hashing variant performs well on the mixed Microsoft dataset. It groups only about 21\% of the benign samples, and interestingly, all detected benign samples are grouped into pure communities (31 in total). This perfect purity is maintained even when changing the similarity threshold hyper-parameter, as will be shown in the next section. {\sf TrapNet} shows modest results with the FH sequence variant on the MS dataset. Similarly, its performance is lower on the TN12k dataset with both FH variants.


\subsection{Similarity Threshold Analysis}
\label{sec_similarity_threshold_analysis}

In this section, we analyze the effectiveness of {\sf TrapNet} while varying the similarity threshold. The threshold is defined as a percentage of the average FloatHash similarity for a given content type. We study the effect of this percentage on the overall performance of {\sf TrapNet} in both malware-only and mixed scenarios.


\subsubsection{Malware-only Scenario}

Here, we track the effect of varying the threshold on the Microsoft and TN12k malware datasets. The effect is measured in terms of coverage, purity, and the number of detected communities.

\begin{scriptsize}
\begin{figure}[H]
    \centering
    \subfloat[\scriptsize Purity \% - Hashing]{%
        \includegraphics[width=0.33\textwidth, trim=0cm 0cm 0cm 0cm, clip]{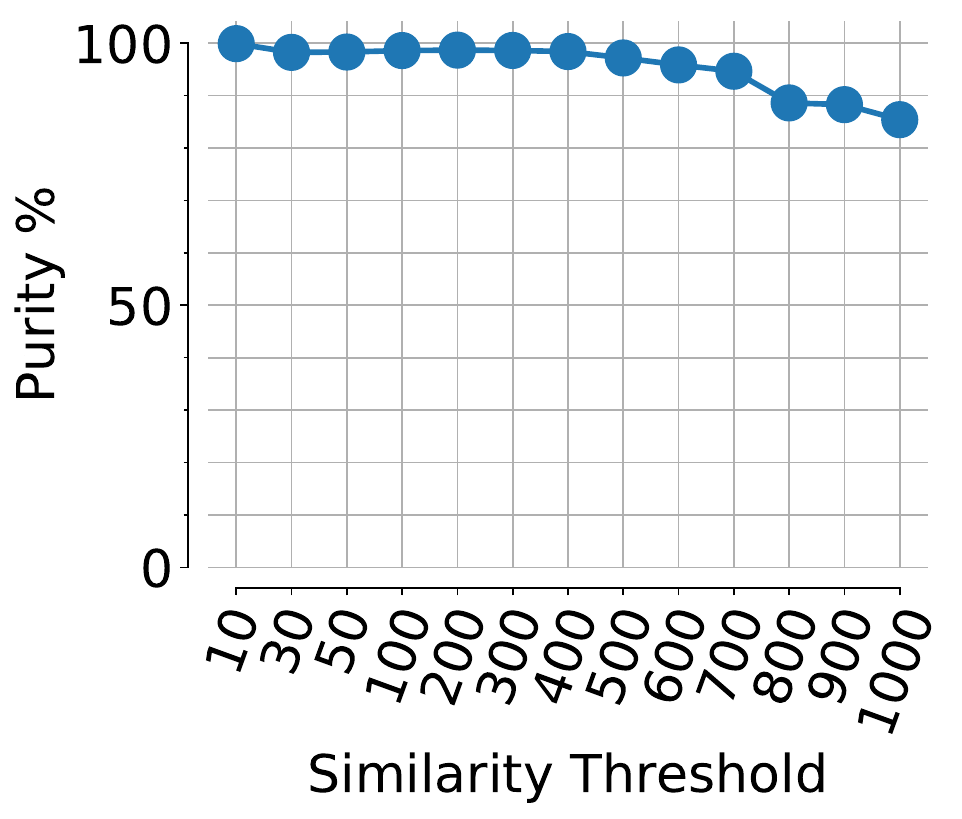}%
        \label{fig_microsoft_hashing_threshold_purity}%
    }
    \hfill
    \subfloat[\scriptsize Coverage \% - Hashing]{%
        \includegraphics[width=0.33\textwidth, trim=0cm 0cm 0cm 0cm, clip]{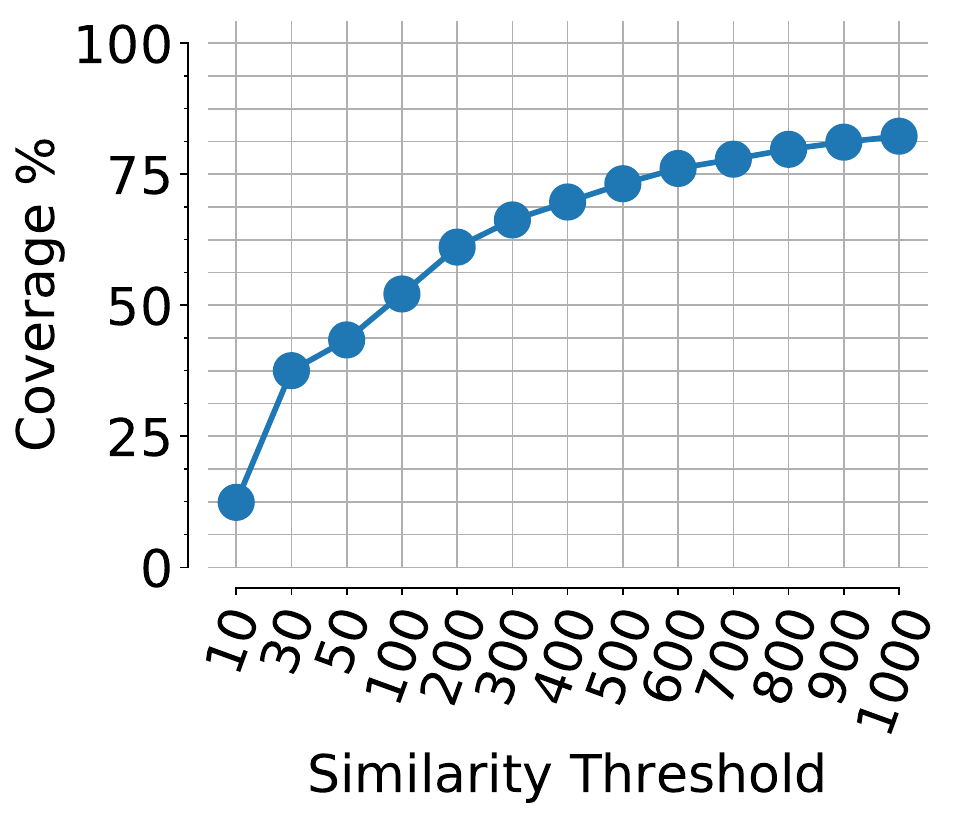}%
        \label{fig_microsoft_hashing_threshold_coverage}%
    }
    \hfill
    \subfloat[\scriptsize \# Community - Hashing]{%
        \includegraphics[width=0.33\textwidth, trim=0cm 0cm 0cm 0cm, clip]{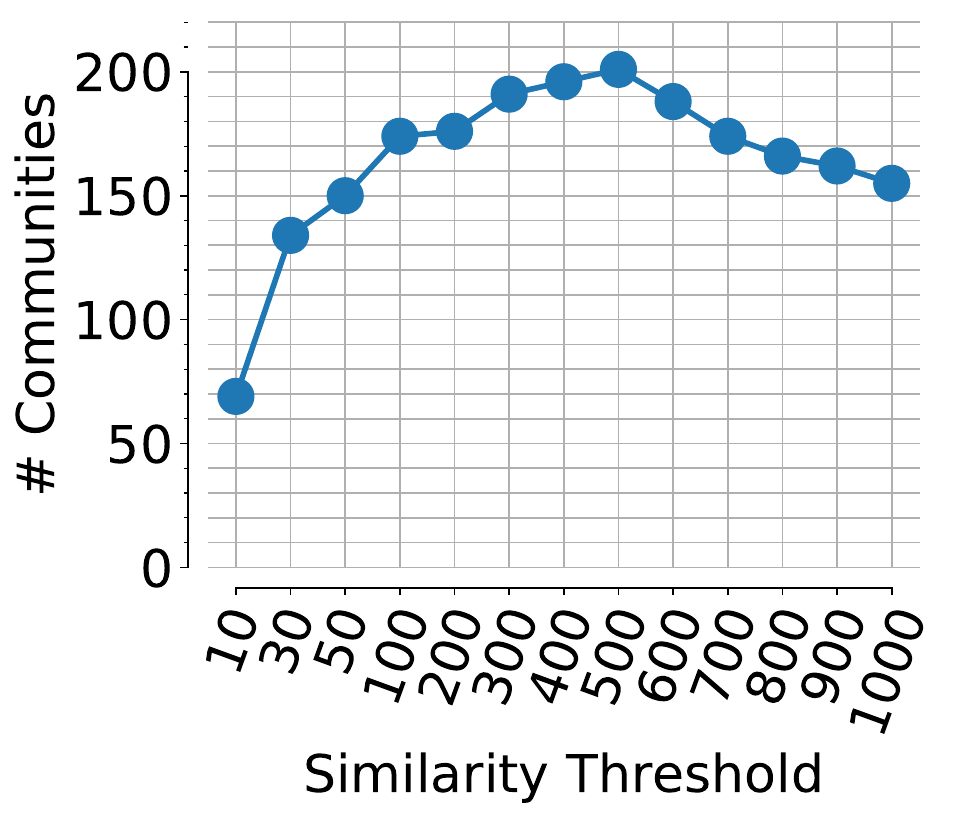}%
        \label{fig_microsoft_hashing_threshold_community}%
    }
    \\
    \subfloat[\scriptsize Purity \% - Sequence]{%
        \includegraphics[width=0.33\textwidth, trim=0cm 0cm 0cm 0cm, clip]{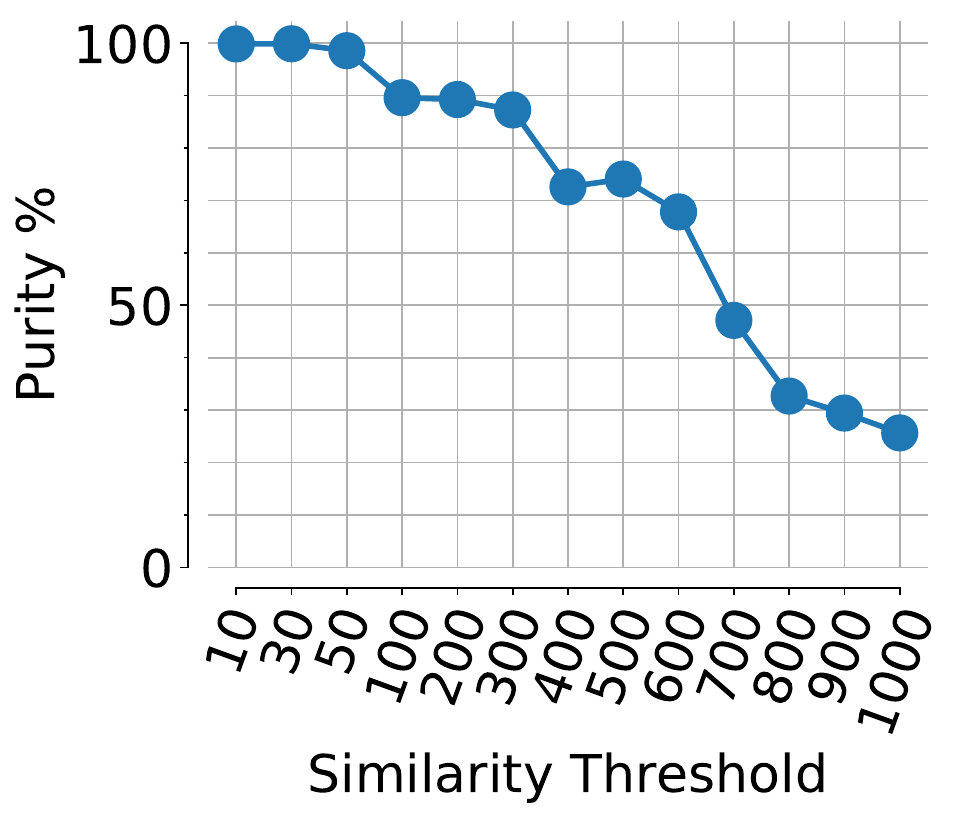}%
        \label{fig_microsoft_sequence_threshold_purity}%
    }
    \hfill
    \subfloat[\scriptsize Coverage \% - Sequence]{%
        \includegraphics[width=0.33\textwidth, trim=0cm 0cm 0cm 0cm, clip]{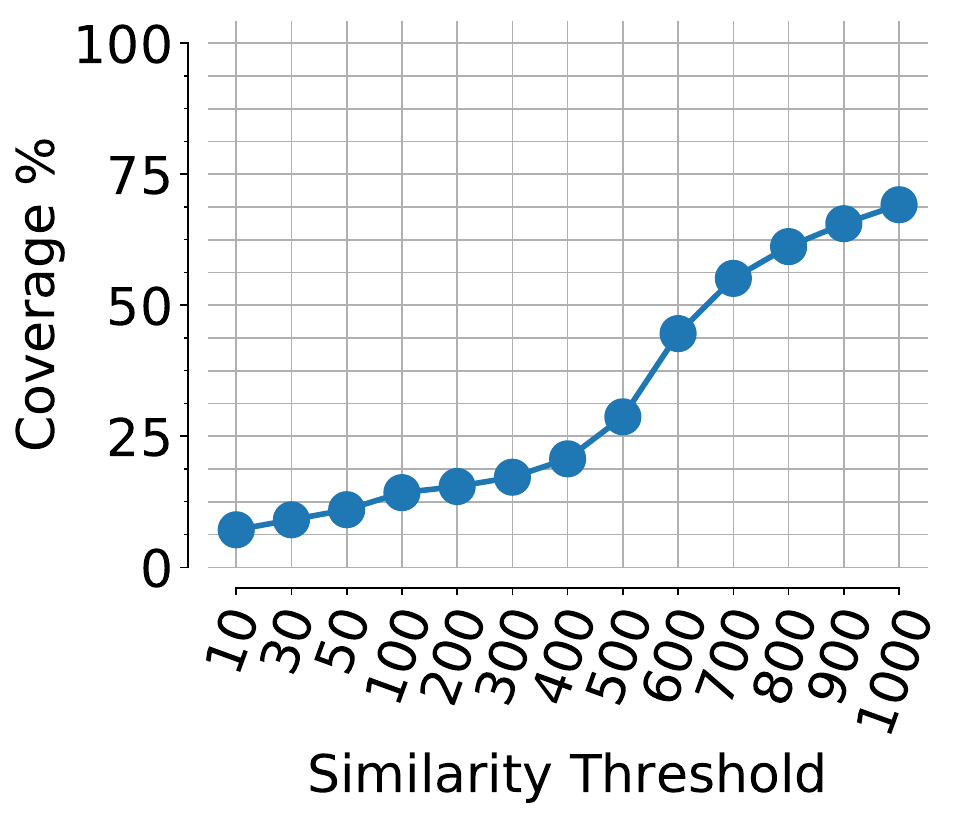}%
        \label{fig_microsoft_sequence_threshold_coverage}%
    }
    \hfill
    \subfloat[\scriptsize \# Community - Sequence]{%
        \includegraphics[width=0.33\textwidth, trim=0cm 0cm 0cm 0cm, clip]{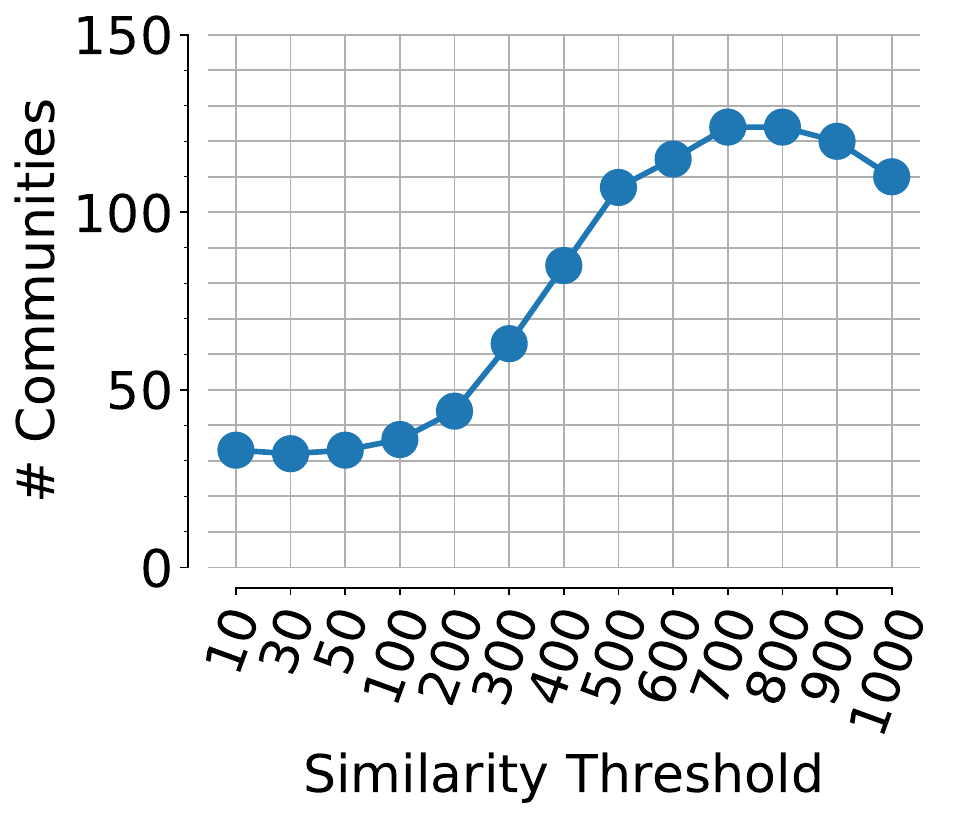}%
        \label{fig_microsoft_sequence_threshold_community}%
    }
    \caption{Similarity Threshold Analysis on Microsoft Dataset}
    \label{fig_microsoft_threshold_hyper}
\end{figure}
\end{scriptsize}


Figure \ref{fig_microsoft_threshold_hyper} shows the results of the threshold analysis on the Microsoft dataset. It plots the change in coverage, purity, and the number of communities over the threshold percentage range (1\% - 100\%). In the FH hashing variant, the threshold has a small effect on purity. However, coverage is highly dependent on the threshold, increasing from 12\% to 82\%. The number of communities increases with the threshold until it peaks at 201 communities with a 50\% threshold, after which it decreases. In the FH sequence variant, the results are different. Purity is highly affected by the threshold; a higher threshold leads to lower purity. Coverage increases slightly from 1\% to 50\% and then ascends rapidly. The change in the number of communities is very similar to the change in coverage. The detailed results for each threshold percentage on the Microsoft dataset are provided in Table \ref{tab_microsoft_malware_effective}, which also details the distribution of communities.

\begin{table}[!htbp]
\centering
\tabalter
\begin{tabular}{lcccc}
\hline \hline
\tabheader
 	           & \textbf{Malware} & \textbf{Malware} \\
\tabheader
\textbf{Threshold \%} & \textbf{Coverage\% / Purity\%} & \textbf{Distribution\footnote{Pure-2Mixed-3Mixed-NMixed} / \#Communities} \\
\hline\hline
\textbf{Hashing Variant} & & \\
\hline
\textit{  1\%}& 12.4\%  / 100.0\% &  (69-0-0-0) / 69   \\
\textit{  3\%}& 37.52\% / 98.0\% &  (126-2-1-5) / 134 \\
\textit{  5\%}& 43.39\% / 98.0\% &  (141-3-1-5) / 150 \\
\textit{ 10\%}& 52.16\% / 99.0\% &  (165-3-1-5) / 174 \\
\textit{ 20\%}& 61.11\% / 99.0\% &  (167-3-1-5) / 176 \\
\textit{ 30\%}& 66.29\% / 99.0\% &  (181-4-1-5) / 191 \\
\textit{ 40\%}& 69.71\% / 98.0\% &  (185-4-2-5) / 196 \\
\textit{ 50\%}& 73.20\% / 97.0\% &  (189-5-2-5) / 201 \\
\textit{ 60\%}& 76.11\% / 96.0\% &  (175-6-2-5) / 188 \\
\textit{ 70\%}& 77.88\% / 95.0\% &  (162-4-3-5) / 174 \\
\textit{ 80\%}& 79.77\% / 89.0\% &  (151-6-4-5) / 166 \\
\textit{ 90\%}& 81.13\% / 88.0\% &  (146-8-2-6) / 162 \\
\textit{100\%}& 82.27\% / 85.0\% &  (138-8-3-6) / 155 \\
\hline
\textbf{Sequence Variant} & & \\
\hline
\textit{  1\%}& 7.16\% / 100.0\% & (33-0-0-0) / 33 \\
\textit{  3\%}& 9.07\% / 100.0\% & (32-0-0-0) / 32 \\
\textit{  5\%}& 11.0\% / 99.0\% &  (32-1-0-0) / 33 \\
\textit{ 10\%}& 14.25\% / 90.0\% & (34-2-0-0) / 36 \\
\textit{ 20\%}& 15.42\% / 89.0\% & (42-0-2-0) / 44 \\
\textit{ 30\%}& 17.2\% / 87.0\% &  (58-3-0-2) / 63 \\
\textit{ 40\%}& 20.7\% / 73.0\% &  (76-4-1-4) / 85 \\
\textit{ 50\%}& 28.72\% / 74.0\% & (95-6-1-5) / 107 \\
\textit{ 60\%}& 44.6\% / 68.0\% &  (96-9-5-5) / 115 \\
\textit{ 70\%}& 55.13\% / 47.0\% & (98-15-3-8) / 124 \\
\textit{ 80\%}& 61.2\% / 33.0\% &  (93-17-3-11) / 124 \\
\textit{ 90\%}& 65.57\% / 29.0\% & (87-11-7-15) / 120 \\
\textit{100\%}& 69.18\% / 26.0\% & (78-9-8-15) / 110 \\
\hline\hline
\end{tabular}
\caption{Malware-only Scenario on Microsoft Dataset (Total time: 10 Seconds)}
\label{tab_microsoft_malware_effective}
\end{table}

\begin{scriptsize}

\begin{figure}[H]
    \centering
    \subfloat[\scriptsize Purity \% - Hashing]{%
        \includegraphics[width=0.33\textwidth, trim=0cm 0cm 0cm 0cm, clip]{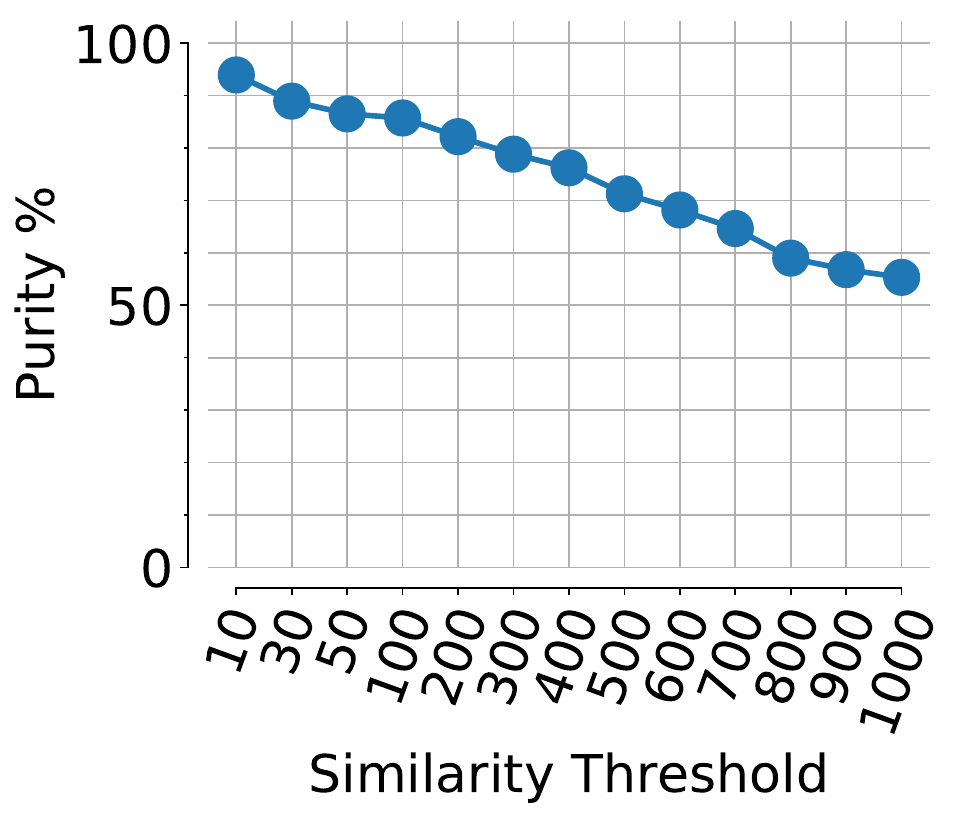}%
        \label{fig_trapnet12k_hashing_threshold_purity}%
    }
    \hfill
    \subfloat[\scriptsize Coverage \% - Hashing]{%
        \includegraphics[width=0.33\textwidth, trim=0cm 0cm 0cm 0cm, clip]{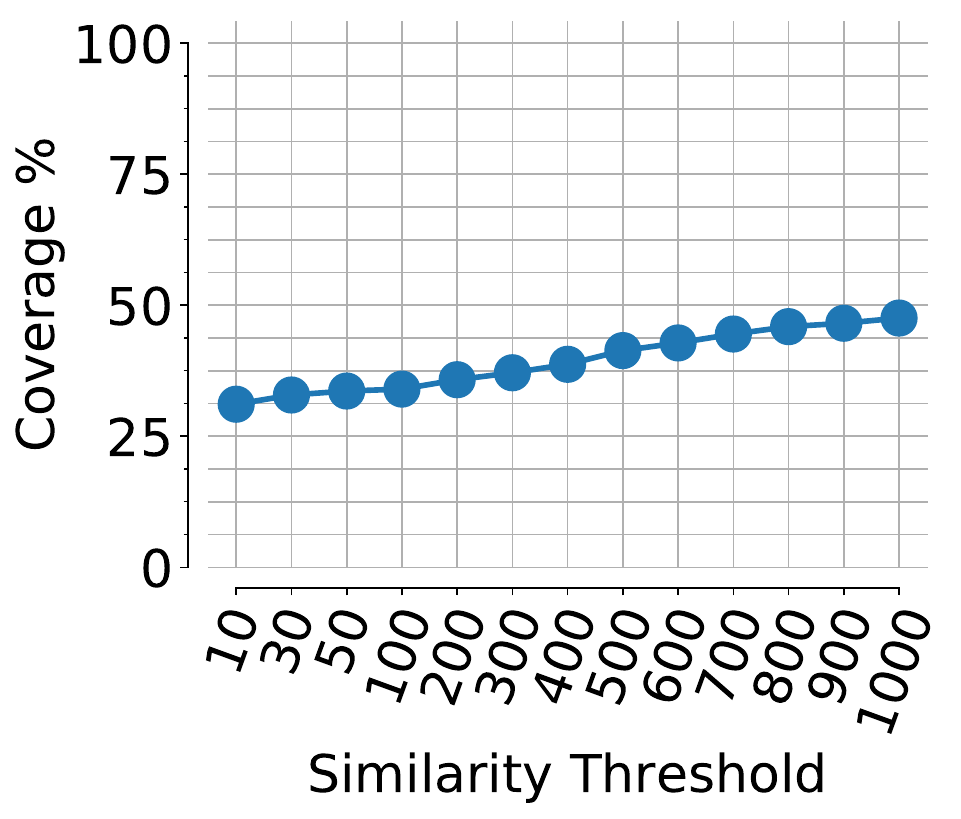}%
        \label{fig_trapnet12k_hashing_threshold_coverage}%
    }
    \hfill
    \subfloat[\scriptsize \# Community - Hashing]{%
        \includegraphics[width=0.33\textwidth, trim=0cm 0cm 0cm 0cm, clip]{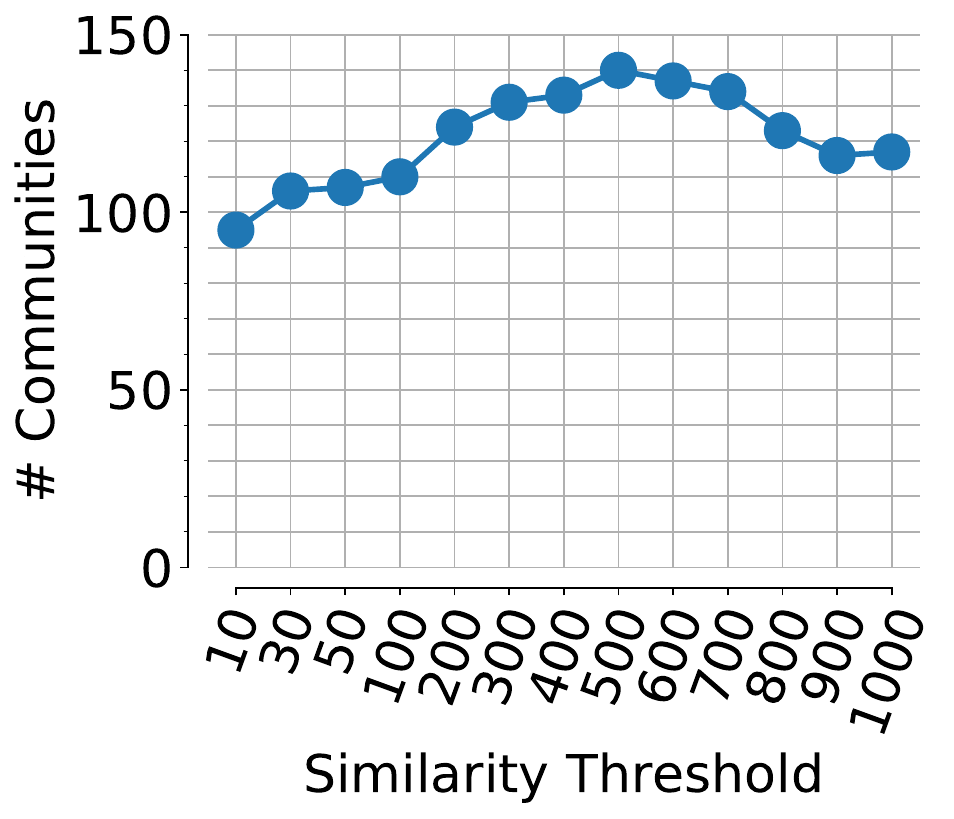}%
        \label{fig_trapnet12k_hashing_threshold_community}%
    }
    \\
    \subfloat[\scriptsize Purity \% - Sequence]{%
        \includegraphics[width=0.33\textwidth, trim=0cm 0cm 0cm 0cm, clip]{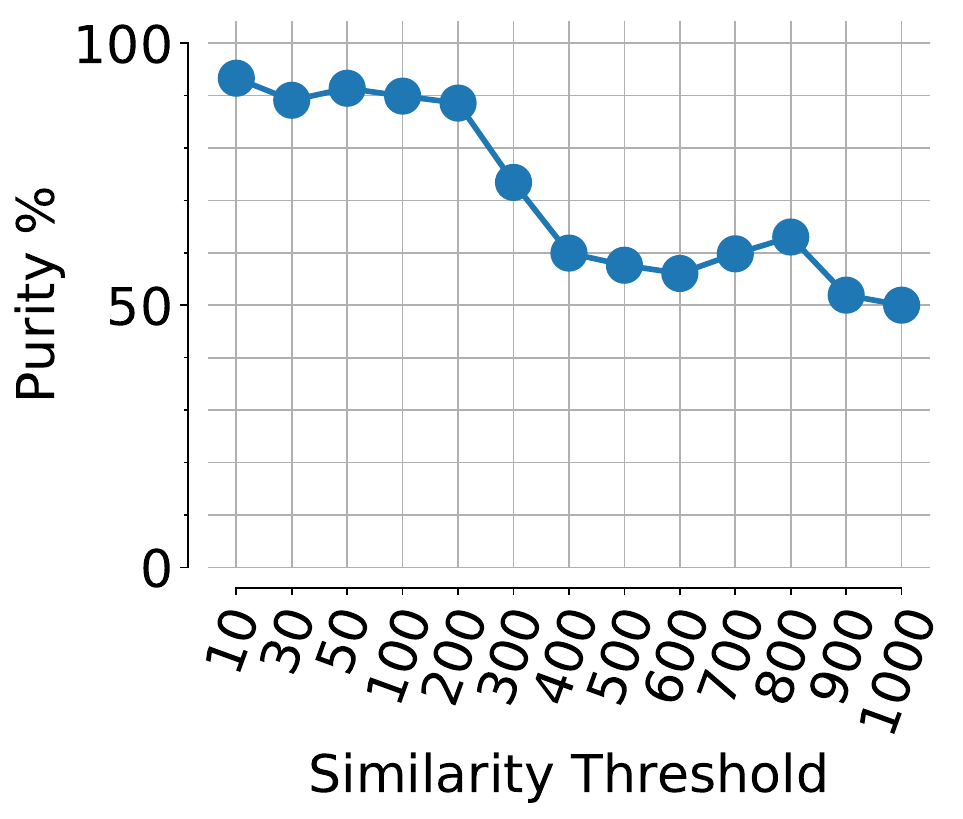}%
        \label{fig_trapnet12k_sequence_threshold_purity}%
    }
    \hfill
    \subfloat[\scriptsize Coverage \% - Sequence]{%
        \includegraphics[width=0.33\textwidth, trim=0cm 0cm 0cm 0cm, clip]{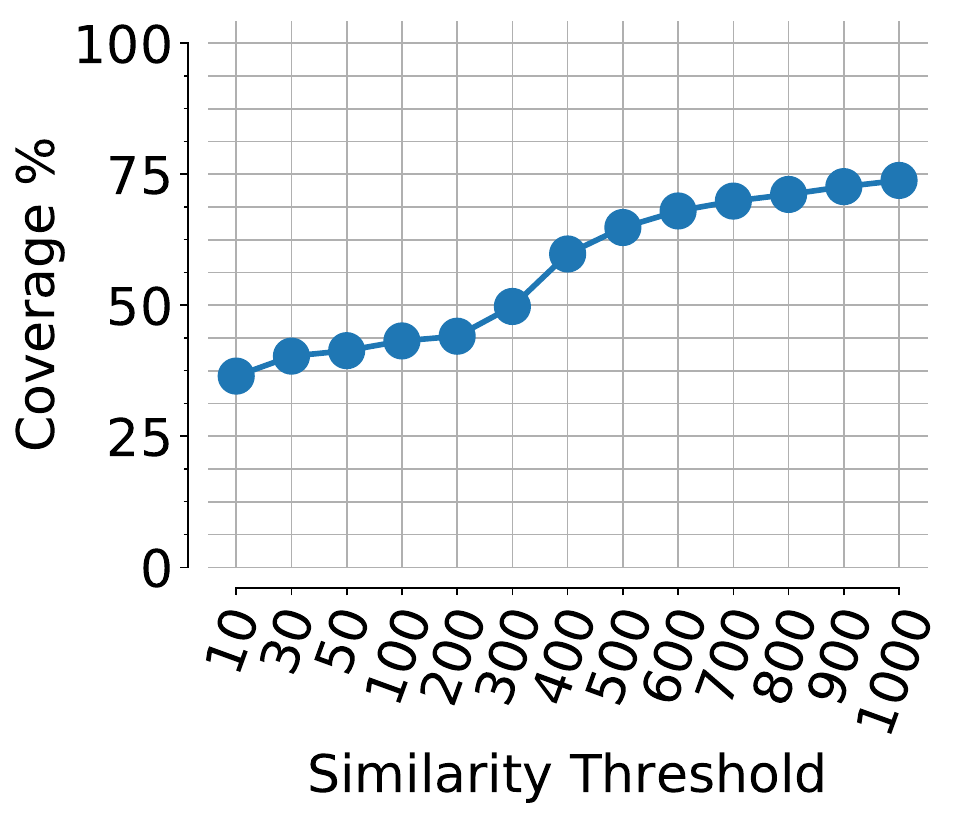}%
        \label{fig_trapnet12k_sequence_threshold_coverage}%
    }
    \hfill
    \subfloat[\scriptsize \# Community - Sequence]{%
        \includegraphics[width=0.33\textwidth, trim=0cm 0cm 0cm 0cm, clip]{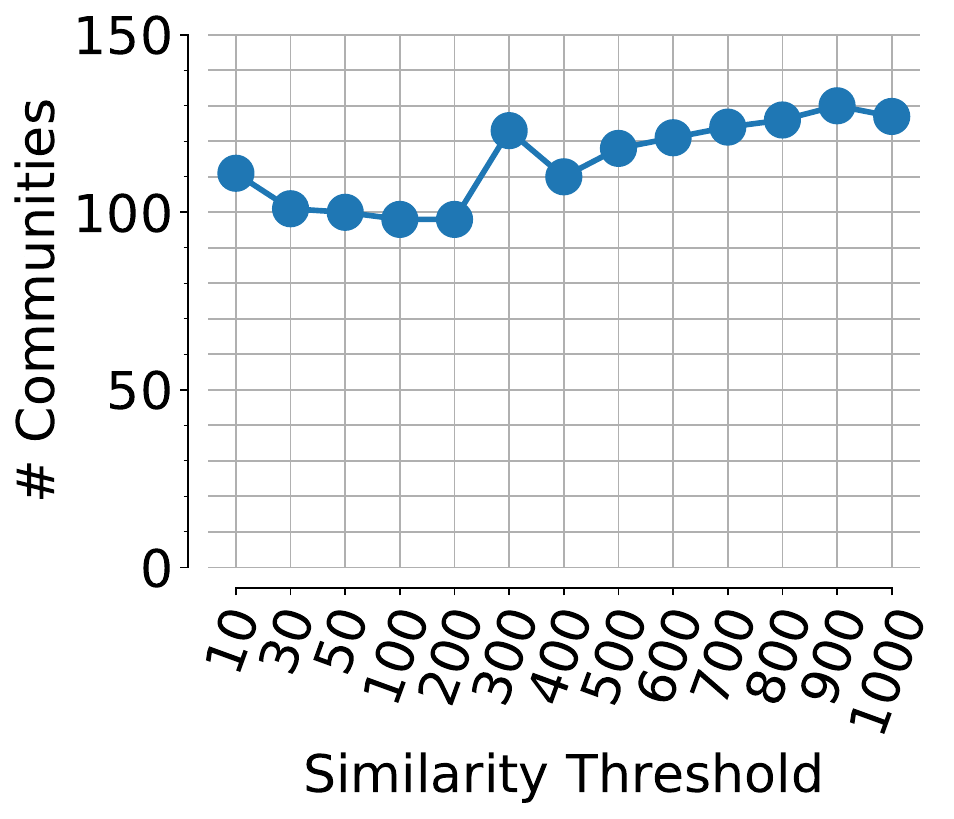}%
        \label{fig_trapnet12k_sequence_threshold_community}%
    }
    \caption{Similarity Threshold Analysis on {\sf TrapNet} Dataset}
    \label{fig_trapnet12k_threshold_hyper}
\end{figure}
\end{scriptsize}


Figure \ref{fig_trapnet12k_threshold_hyper} depicts the effects of changing the threshold percentage on the TN12k malware dataset. In contrast to Figure \ref{fig_microsoft_threshold_hyper}, the effects for the FH hashing and sequence variants are very similar. Moreover, the results for coverage, purity, and the number of communities are comparable. The detailed results for the TN12k dataset are depicted in Table \ref{tab_trapnet12k_malware_effective}. It is important to mention that each threshold experiment shown in Table \ref{tab_microsoft_malware_effective} and Table \ref{tab_trapnet12k_malware_effective} took only ten seconds, including both similarity computation and community detection.


\begin{table}[!htbp]
\centering
\tabalter
\begin{tabular}{lcccc}
\hline \hline
\tabheader
 	           & \textbf{Malware} & \textbf{Malware} \\
\tabheader
\textbf{Threshold \%} & \textbf{Coverage\% / Purity\%} & \textbf{Distribution\footnote{Pure-2Mixed-3Mixed-NMixed} / \#Communities} \\
\hline\hline
\textbf{Hashing Variant} & & \\
\hline
\textit{  1\%}& 31.13\% / 94.0\% & (81-9-4-1) / 95     \\
\textit{  3\%}& 32.88\% / 89.0\% & (81-12-6-7) / 106   \\
\textit{  5\%}& 33.64\% / 87.0\% & (77-13-9-8) / 107   \\
\textit{ 10\%}& 34.01\% / 86.0\% & (77-15-7-11) / 110  \\
\textit{ 20\%}& 35.80\% / 82.0\% & (80-20-8-16) / 124  \\
\textit{ 30\%}& 37.12\% / 79.0\% & (81-23-9-18) / 131  \\
\textit{ 40\%}& 38.74\% / 76.0\% & (82-22-9-20) / 133  \\
\textit{ 50\%}& 41.36\% / 71.0\% & (80-25-11-24) / 140 \\
\textit{ 60\%}& 42.82\% / 68.0\% & (76-23-14-24) / 137 \\
\textit{ 70\%}& 44.52\% / 65.0\% & (73-19-15-27) / 134 \\
\textit{ 80\%}& 45.92\% / 59.0\% & (65-20-12-26) / 123 \\
\textit{ 90\%}& 46.58\% / 57.0\% & (63-15-13-25) / 116 \\
\textit{100\%}& 47.57\% / 55.0\% & (61-16-11-29) / 117 \\
\hline
\textbf{Sequence Variant} & & \\
\hline
\textit{  1\%}& 36.49\% / 93.0\% & (100-7-3-1) / 111   \\
\textit{  3\%}& 40.30\% / 89.0\% & (89-7-4-1) / 101    \\
\textit{  5\%}& 41.34\% / 91.0\% & (88-7-3-2) / 100    \\
\textit{ 10\%}& 43.20\% / 90.0\% & (83-7-5-3) / 98     \\
\textit{ 20\%}& 44.10\% / 89.0\% & (80-8-7-3) / 98     \\
\textit{ 30\%}& 49.78\% / 73.0\% & (101-9-8-5) / 123   \\
\textit{ 40\%}& 59.79\% / 60.0\% & (85-11-8-6) / 110   \\
\textit{ 50\%}& 64.84\% / 58.0\% & (90-9-10-9) / 118   \\
\textit{ 60\%}& 68.00\% / 56.0\% & (91-13-9-8) / 121   \\
\textit{ 70\%}& 69.88\% / 60.0\% & (92-13-11-8) / 124  \\
\textit{ 80\%}& 71.15\% / 63.0\% & (91-15-11-9) / 126  \\
\textit{ 90\%}& 72.64\% / 52.0\% & (92-15-13-10) / 130 \\
\textit{100\%}& 73.82\% / 50.0\% & (87-17-13-10) / 127 \\
\hline\hline
\end{tabular}
\caption{Malware-only Scenario on 12k {\sf TrapNet} samples (Total time: 10 Seconds)}
\label{tab_trapnet12k_malware_effective}
\end{table}


\subsubsection{Mixed Scenario}

Here, we present the effect of the threshold percentage hyper-parameter on benign sample misdetection. The results for malware detection are nearly identical to those of the malware-only scenario; therefore, we focus on the benign detection performance.


\begin{table}[!htbp]
\centering
\tabalter
\begin{tabular}{lcccc}
\hline \hline
\tabheader
 	           & \textbf{Benign} & \textbf{Benign} \\
\tabheader
\textbf{Threshold \%} & \textbf{Coverage\% / Purity\%} & \textbf{Distribution\footnote{Pure-2Mixed-3Mixed-NMixed} / \#Communities} \\
\hline\hline
\textbf{Hashing Variant} & & \\
\hline
\textit{1\%} &  4.6\% / 100.0\% & 	(7-0-0-0) / 7   \\
\textit{3\%} &  5.21\% / 100.0\% & 	(8-0-0-0) / 8   \\
\textit{5\%} &  5.21\% / 100.0\% & 	(7-0-0-0) / 7   \\
\textit{10\%} &  7.58\% / 100.0\% & 	(13-0-0-0) / 13 \\
\textit{20\%} &  9.72\% / 100.0\% & 	(15-0-0-0) / 15 \\
\textit{30\%} &  10.88\% / 100.0\% & 	(17-0-0-0) / 17 \\
\textit{40\%} &  11.86\% / 100.0\% & 	(18-0-0-0) / 18 \\
\textit{50\%} &  13.16\% / 100.0\% & 	(18-0-0-0) / 18 \\
\textit{60\%} &  13.86\% / 100.0\% & 	(19-0-0-0) / 19 \\
\textit{70\%} &  15.26\% / 100.0\% & 	(21-0-0-0) / 21 \\
\textit{80\%} &  17.3\% / 100.0\% & 	(26-0-0-0) / 26 \\
\textit{90\%} &  19.4\% / 100.0\% & 	(30-0-0-0) / 30 \\
\textit{100\%} &  20.93\% / 100.0\% & 	(31-0-0-0) / 31 \\
\hline
\textbf{Sequence Variant} & & \\
\hline
\textit{1\%} &  7.91\% / 84.0\% & 	(7-0-0-1) / 8  \\
\textit{3\%} &  9.86\% / 84.0\% & 	(9-0-0-1) / 10 \\
\textit{5\%} & 10.51\% / 85.0\% & 	(8-0-0-1) / 9  \\
\textit{10\%} & 10.98\% / 86.0\% & 	(7-0-0-1) / 8  \\
\textit{20\%} & 11.86\% / 50.0\% & 	(7-1-0-1) / 9  \\
\textit{30\%} & 15.35\% / 42.0\% & 	(8-2-1-2) / 13 \\
\textit{40\%} & 21.86\% / 32.0\% & 	(9-2-2-3) / 16 \\
\textit{50\%} & 28.37\% / 19.0\% & 	(5-2-3-4) / 14 \\
\textit{60\%} & 33.77\% / 18.0\% & 	(6-2-4-5) / 17 \\
\textit{70\%} & 38.60\% / 15.0\% & 	(6-1-4-8) / 19 \\
\textit{80\%} & 45.21\% / 14.0\% & 	(7-3-5-7) / 22 \\
\textit{90\%} & 51.72\% / 12.0\% & 	(7-9-6-9) / 31 \\
\textit{100\%} & 59.49\% / 5.0\% &      (8-9-9-11) / 37 \\
\hline\hline
\end{tabular}
\caption{Mixed Scenario on Microsoft + Benign Datasets}
\label{tab_microsoft_mixed_effective}
\end{table}


Starting with the mixed Microsoft dataset (10.9k malware + 2.2k benign samples), Table \ref{tab_microsoft_mixed_effective} details the detection results for benign samples in both FH hashing and sequence variants. In the FH hashing variant, {\sf TrapNet} achieves excellent results with low coverage for benign samples. Furthermore, all detected benign samples are part of pure communities, and this perfect purity is maintained across the entire threshold range. In the FH sequence variant, {\sf TrapNet} displays lower performance, with considerable coverage and low purity over most of the threshold range.


\begin{table}[!htbp]
\centering
\tabalter
\begin{tabular}{lcccc}
\hline \hline
\tabheader
 	           & \textbf{Benign} & \textbf{Benign} \\
\tabheader
\textbf{Threshold \%} & \textbf{Coverage\% / Purity\%} & \textbf{Distribution\footnote{Pure-2Mixed-3Mixed-NMixed} / \#Communities} \\
\hline\hline
\textbf{Hashing Variant} & & \\
\hline
\textit{1\%} & 24.13\% / 92.0\% & 	  (8-0-0-2) / 10 \\
\textit{3\%} & 25.47\% / 86.0\% & 	  (9-1-0-2) / 12 \\
\textit{5\%} & 26.13\% / 84.0\% & 	 (10-4-1-2) / 17 \\
\textit{10\%} & 26.79\% / 83.0\% & 	 (12-4-1-3) / 20 \\
\textit{20\%} & 28.12\% / 71.0\% & 	 (14-3-2-6) / 25 \\
\textit{30\%} & 28.98\% / 68.0\% & 	 (21-4-2-7) / 34 \\
\textit{40\%} & 30.03\% / 65.0\% & 	 (18-3-4-8) / 33 \\
\textit{50\%} & 31.96\% / 62.0\% & 	(16-6-3-10) / 35 \\
\textit{60\%} & 33.91\% / 59.0\% & 	(14-7-3-12) / 36 \\
\textit{70\%} & 35.44\% / 55.0\% & 	(11-4-7-15) / 37 \\
\textit{80\%} & 37.03\% / 52.0\% & 	(12-5-3-15) / 35 \\
\textit{90\%} & 37.97\% / 46.0\% & 	(13-4-3-14) / 34 \\
\textit{100\%} & 38.57\% / 44.0\% & 	(12-4-3-12) / 31 \\
\hline
\textbf{Sequence Variant} & & \\
\hline
\textit{1\%} & 38.24\% / 97.0\% & 	 (5-0-0-0) / 5  \\
\textit{3\%} & 40.17\% / 93.0\% & 	 (6-0-0-0) / 6  \\
\textit{5\%} & 40.81\% / 74.0\% & 	 (6-0-0-0) / 6  \\
\textit{10\%} & 41.92\% / 74.0\% & 	 (6-1-0-1) / 8  \\
\textit{20\%} & 59.54\% / 69.0\% & 	 (7-1-0-2) / 10 \\
\textit{30\%} & 65.59\% / 67.0\% & 	 (8-0-1-2) / 11 \\
\textit{40\%} & 67.37\% / 59.0\% & 	 (8-0-1-2) / 11 \\
\textit{50\%} & 68.25\% / 51.0\% & 	 (8-0-1-2) / 11 \\
\textit{60\%} & 69.39\% / 50.0\% & 	 (8-0-1-2) / 11 \\
\textit{70\%} & 70.63\% / 54.0\% & 	 (9-1-1-2) / 13 \\
\textit{80\%} & 71.54\% / 48.0\% & 	(11-0-0-3) / 14 \\
\textit{90\%} & 72.44\% / 48.0\% & 	(10-0-0-3) / 13 \\
\textit{100\%} & 73.37\% / 47.0\% & 	(10-0-1-3) / 14 \\
\hline\hline
\end{tabular}
\caption{Mixed Scenario on 12k {\sf TrapNet} + Benign Dataset}
\label{tab_trapnet12k_mixed_effective}
\end{table}


Table \ref{tab_trapnet12k_mixed_effective} presents the results of the threshold analysis on the mixed TN12k dataset. The performance is lower compared to the Microsoft dataset (Table \ref{tab_microsoft_mixed_effective}). Overall, the FH hashing variant outperforms the FH sequence variant. The dynamics of coverage and purity are similar in both variants: a smaller threshold percentage leads to higher purity and lower coverage, and vice versa.


\section{Scalability Evaluation}
\label{sec_efficiency_evaluation}

In this section, we analyze the scalability of {\sf TrapNet} by answering the following questions: How does scalability influence the overall performance for different FH variants? How does the runtime increase with dataset size? How quickly can {\sf TrapNet} handle a day's worth of malware samples (250k samples \cite{web_malware_variant_2018})? In our runtime analysis, similar to MutantX-S \cite{Hu13MutantX}, we consider the clustering runtime starting from the FH digests, which includes building the malware similarity network and executing community detection.


\begin{scriptsize}
\begin{figure}[H]
    \centering
    \subfloat[\scriptsize Purity \% - Hashing]{%
        \includegraphics[width=0.33\textwidth, trim=0cm 0cm 0cm 0.0cm, clip]{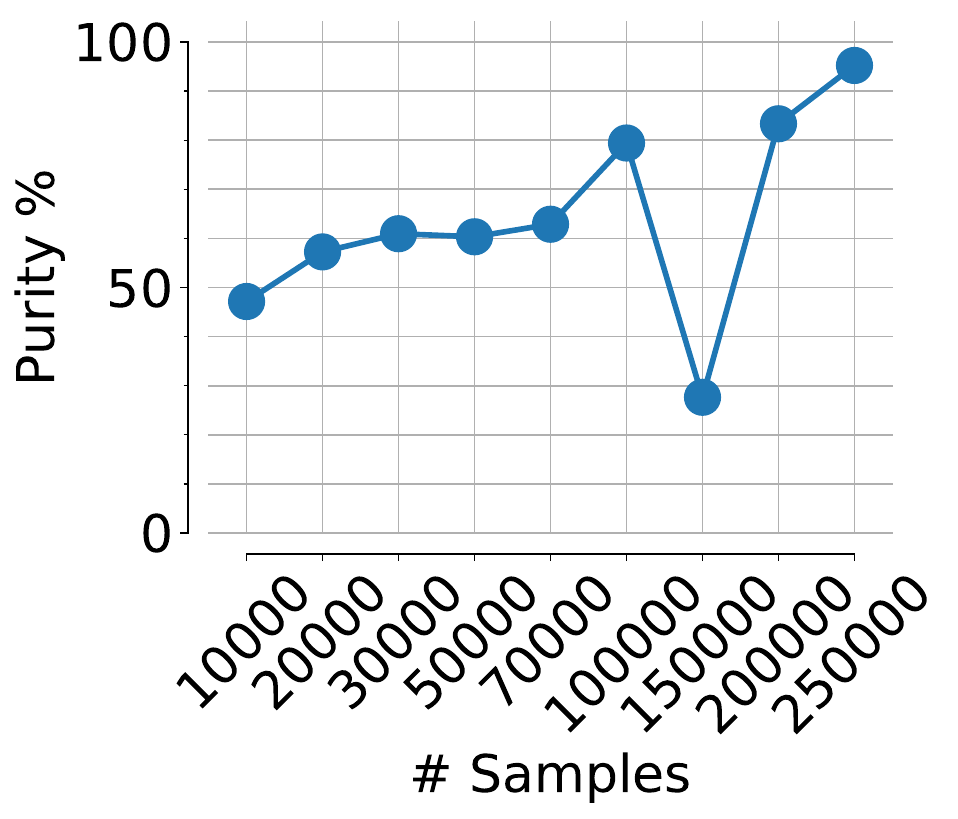}%
        \label{fig_trapnet250k_hashing_scalability_purity}%
    }
    \hfill
    \subfloat[\scriptsize Coverage \% - Hashing]{%
        \includegraphics[width=0.33\textwidth, trim=0cm 0cm 0cm 0.0cm, clip]{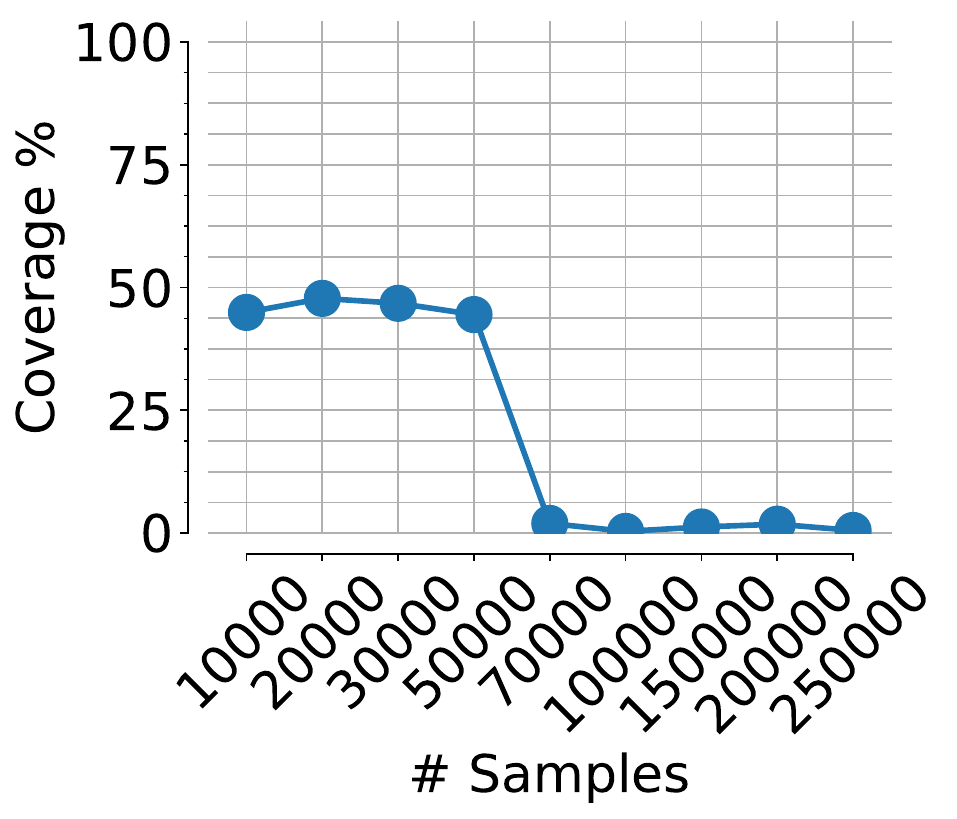}%
        \label{fig_trapnet250k_hashing_scalability_coverage}%
    }
    \hfill
    \subfloat[\scriptsize \# Community - Hashing]{%
        \includegraphics[width=0.33\textwidth, trim=0cm 0cm 0cm 0.0cm, clip]{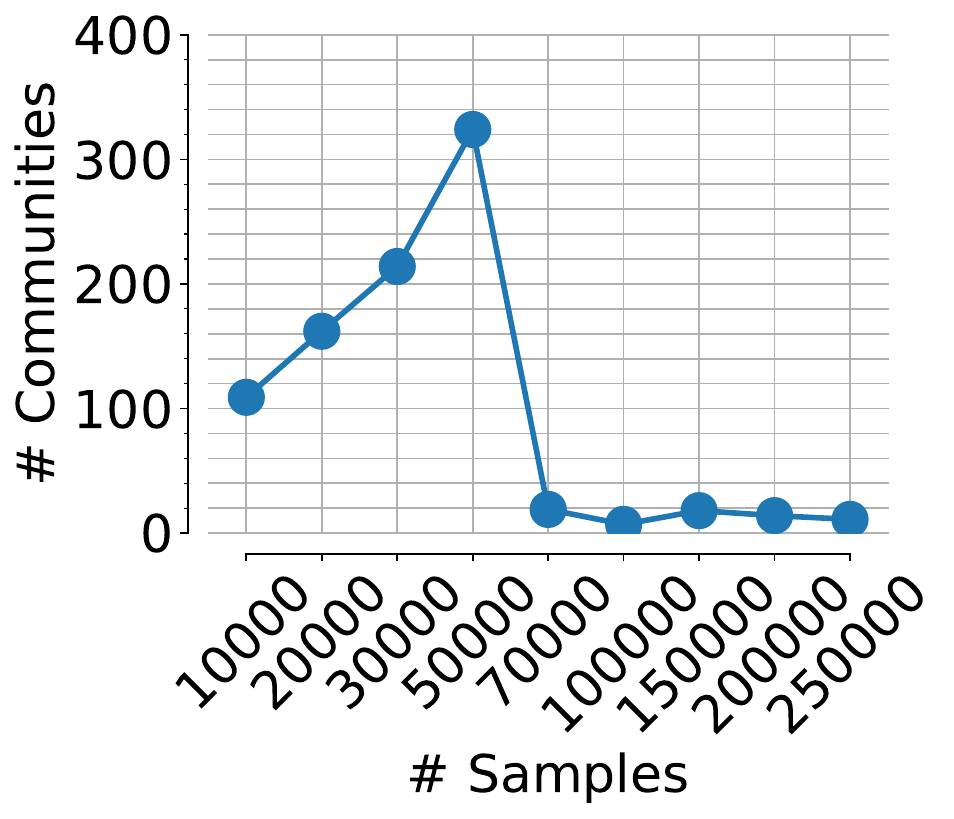}%
        \label{fig_trapnet250k_hashing_scalability_community}%
    }
    \\
    \subfloat[\scriptsize Purity \% - Sequence]{%
        \includegraphics[width=0.33\textwidth, trim=0cm 0cm 0cm 0cm, clip]{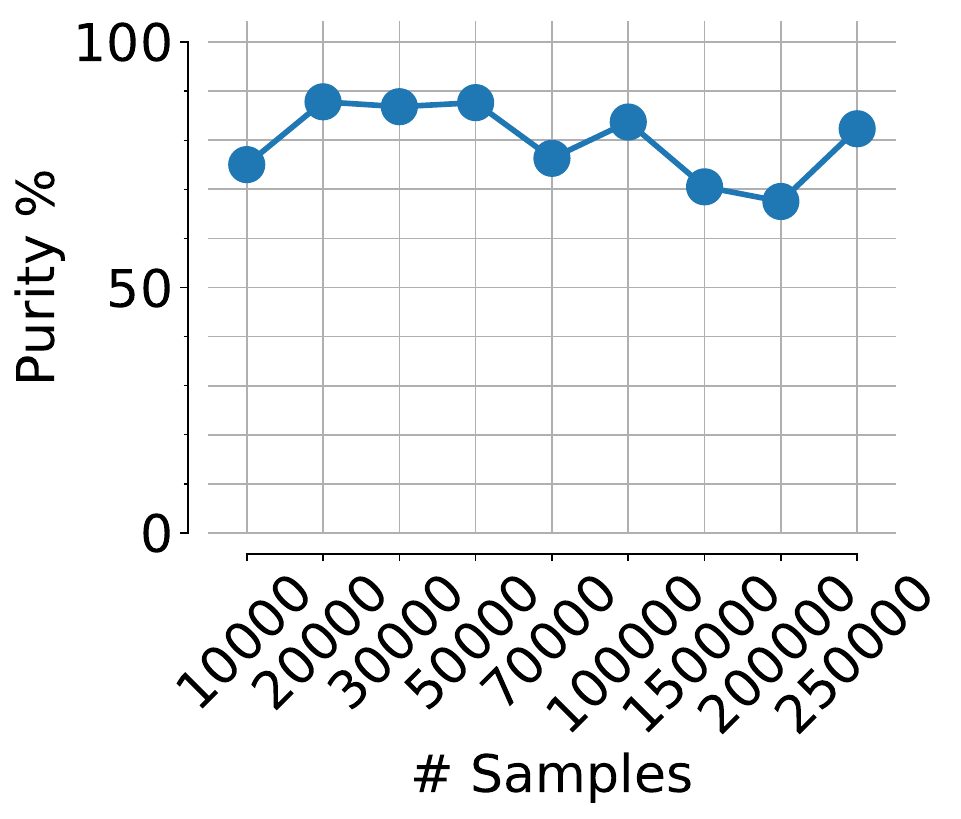}%
        \label{fig_trapnet250k_sequence_scalability_purity}%
    }
    \hfill
    \subfloat[\scriptsize Coverage \% - Sequence]{%
        \includegraphics[width=0.33\textwidth, trim=0cm 0cm 0cm 0cm, clip]{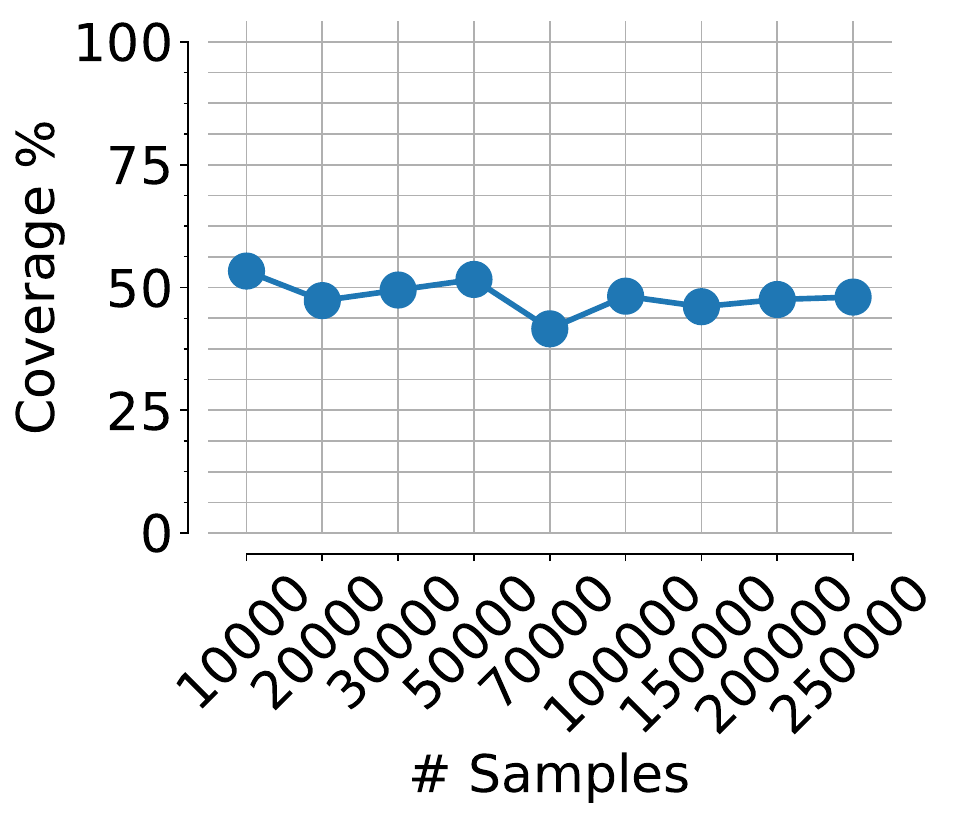}%
        \label{fig_trapnet250k_sequence_scalability_coverage}%
    }
    \hfill
    \subfloat[\scriptsize \# Community - Sequence]{%
        \includegraphics[width=0.33\textwidth, trim=0cm 0cm 0cm 0cm, clip]{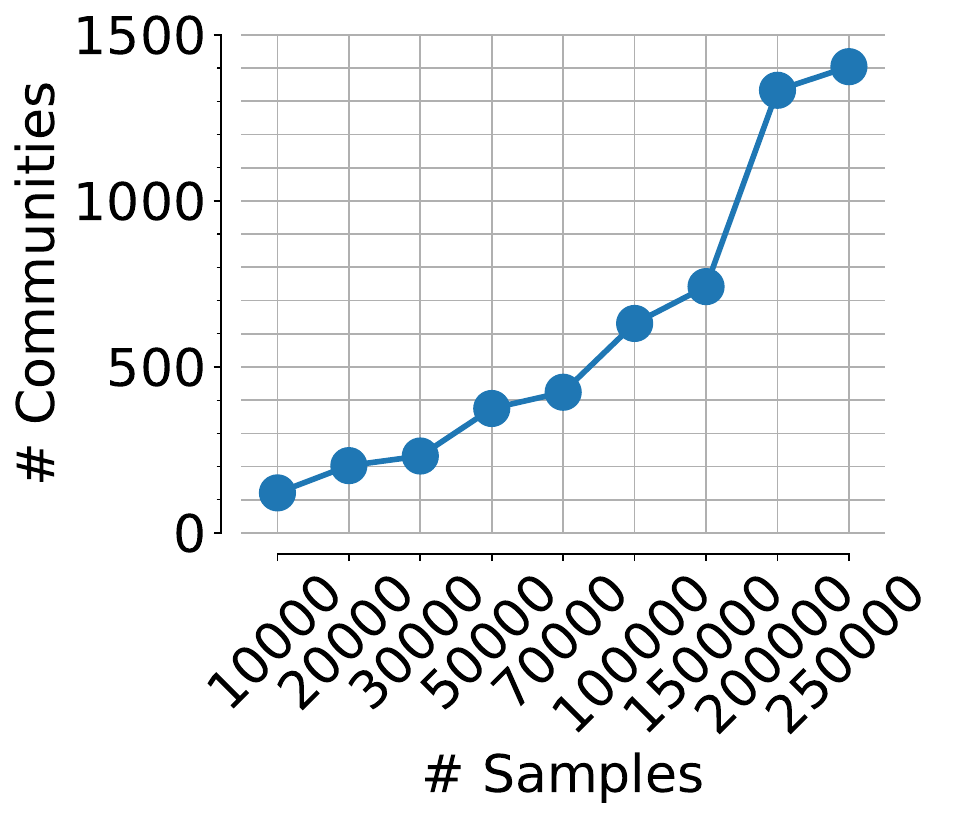}%
        \label{fig_trapnet250k_sequence_scalability_community}%
    }
    \caption{Scalability Analysis on {\sf TrapNet} Dataset}
    \label{fig_trapnet250k_scalability_hyper}
\end{figure}
\end{scriptsize}


Figure \ref{fig_trapnet250k_scalability_hyper} presents the coverage, purity, and number of detected communities for different {\sf TrapNet} dataset sizes, ranging from 10k to 250k malware samples. For this analysis, the similarity threshold for each FH variant was set to the optimal value determined from the TN12k experiments. The FH hashing variant shows relatively good coverage and purity for dataset sizes from 10k to 50k. Surprisingly, the coverage drops sharply for the 70k sample dataset and remains low for all larger datasets. Consequently, the number of detected communities also drops. From this analysis, we conclude that the FH hashing variant is not scalable, as it gives poor results on large datasets (above 70k samples in our evaluation). The purity percentage appears to increase for larger datasets only because the small number of detected malware samples are grouped into very few communities.


\begin{table}[!htbp]
\centering
\tabalter
\begin{tabular}{lcccc}
\hline \hline
\tabheader
 	           & \textbf{Malware} & \textbf{Malware} \\
\tabheader
\textbf{\#Samples} & \textbf{Coverage\% / Purity\%} & \textbf{Distribution\footnote{Pure-2Mixed-3Mixed-NMixed} / \#Communities} \\
\hline\hline
\textbf{Hashing Variant} & & \\
\hline
\textit{10k }& 44.93\% / 47.0\% & 	(57-17-8-27) / 109 \\
\textit{20k }& 47.78\% / 57.0\% & 	(87-25-16-34) / 162 \\
\textit{30k }& 46.78\% / 61.0\% & 	(113-36-14-51 )/ 214 \\
\textit{50k }& 44.49\% / 60.0\% & 	(155-70-35-64) / 324 \\
\textit{70k }&  1.96\% / 63.0\% & 	(6-2-2-9) / 19 \\
\textit{100k}&  0.34\% / 79.0\% & 	(5-1-1-0) / 7 \\
\textit{150k}&  1.24\% / 28.0\% & 	(7-1-3-7) / 18 \\
\textit{200k}&  1.81\% / 83.0\% & 	(7-2-0-5) / 14 \\
\textit{250k}&  0.53\% / 95.0\% & 	(7-1-2-1) / 11 \\
\hline
\textbf{Sequence Variant} & & \\
\hline
\textit{10k }& 	53.35\% / 75.0\% & 	    (103-10-4-4) / 121  \\
\textit{20k }& 	47.33\% / 88.0\% & 	   (167-17-6-13) / 203  \\
\textit{30k }& 	49.49\% / 87.0\% & 	   (180-30-9-13) / 232  \\
\textit{50k }& 	51.64\% / 88.0\% & 	  (279-55-20-21) / 375  \\
\textit{70k }& 	41.59\% / 76.0\% & 	  (293-71-31-29) / 424  \\
\textit{100k}& 	48.24\% / 84.0\% & 	 (431-114-47-39) / 631  \\
\textit{150k}& 	46.09\% / 70.0\% & 	 (427-170-82-63) / 742  \\
\textit{200k}& 	47.56\% / 68.0\% & 	(847-258-133-95) / 1333 \\
\textit{250k}& 	48.05\% / 82.0\% & 	(946-256-105-97) / 1404 \\
\hline\hline
\end{tabular}
\caption{Scalability Analysis on 250k {\sf TrapNet} Dataset}
\label{tab_trapnet250k_scalability}
\end{table}


On the other hand, the FH sequence variant demonstrates surprisingly good and stable results across all dataset sizes. The overall coverage is approximately 50\% in all configurations. Even more interesting is the achieved purity, which is 80\% on average. The number of detected communities increases with the dataset size. Overall, the sequence variant shows scalable and stable characteristics compared to the FH hashing variant. Moreover, its performance tends to be higher on the {\sf TrapNet} dataset, which we believe is more realistic (Section \ref{sec_dataset}). The detailed scalability results are presented in Table \ref{tab_trapnet250k_scalability}.


\begin{center}
\begin{figure}[H]
    \includegraphics[width=0.85\textwidth, trim=0cm 0cm 0cm 0cm, clip]{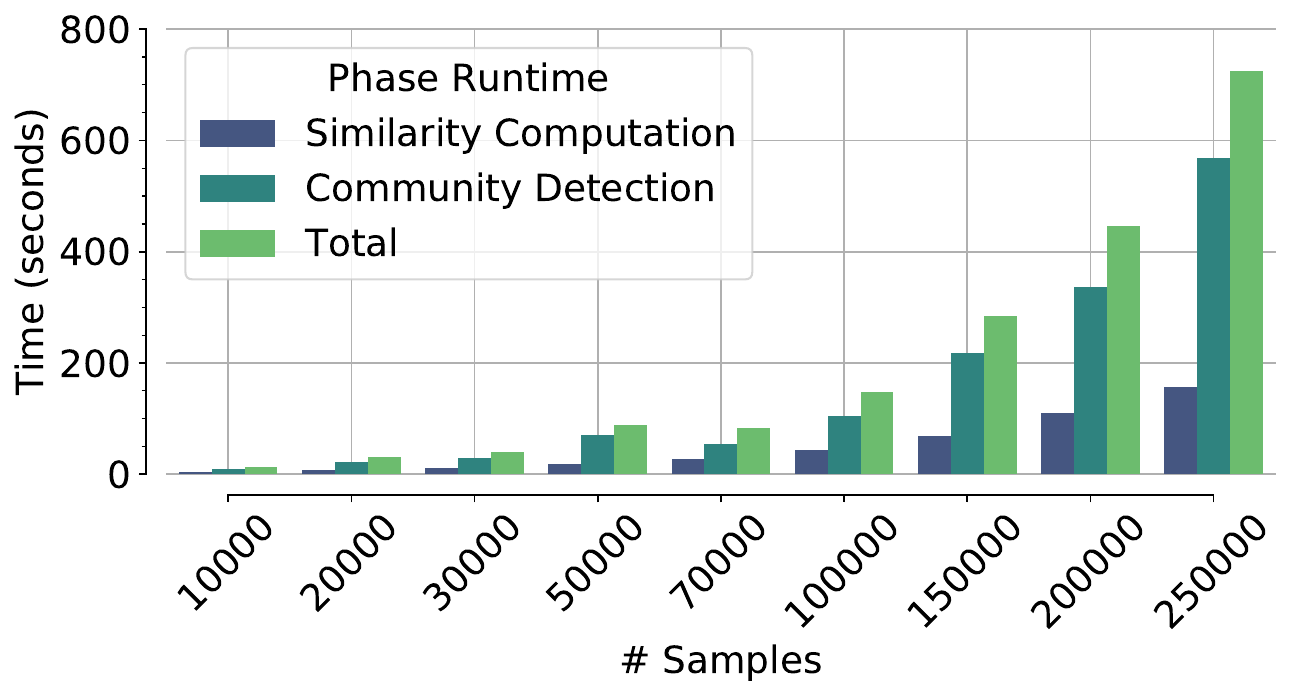}
    \caption{Runtime vs. Dataset Size}
   \label{fig_runtime_dataset_size}
\end{figure}
\end{center}


Figure \ref{fig_runtime_dataset_size} depicts the runtime of the {\sf TrapNet} framework on different dataset sizes. The runtime is nearly identical for both FH variants. The most noticeable feature of {\sf TrapNet} is its rapid malware clustering. {\sf TrapNet} takes only 10 seconds to cluster 10k malware samples with good results, as presented previously. {\sf TrapNet} is significantly faster than other solutions, such as MutantX-S \cite{Hu13MutantX}, which clustered about 130k malware samples in approximately 90 minutes with 82\% precision (a metric similar to our purity). In contrast, {\sf TrapNet} with the FH sequence variant processed 250k malware samples from 73 families and achieved 82\% purity in only 12 minutes. This makes {\sf TrapNet} approximately 15 times faster than MutantX-S.

\section{Related Work}
\label{sec_related_work}

Malware fingerprinting is a critical task in cybersecurity due to its potential to prevent or mitigate harmful effects on computer systems. Many solutions have been proposed to fingerprint malware by automatically clustering and classifying it based on static and dynamic features. The solutions in \cite{Gheorghescu05AUTOMATED, Bailey07Automated, Bayer09Scalable} rely on behavioral features to cluster malware, while \cite{Rieck08Learning} leverages dynamic features to classify new malware samples. These solutions share a reliance on runtime and dynamic analysis. In contrast, our work relies exclusively on static features for large-scale malware clustering.

A substantial body of research has explored static analysis as a foundation for scalable malware fingerprinting, particularly in the Android ecosystem. Karbab et al. \cite{karbab2016Fingerprinting} proposed a fuzzy hashing-based fingerprinting approach for Android APK packaging that captures both binary features and the underlying structural properties of the application, achieving 95\% precision in malware detection and family attribution through their ROAR system. Building on this line of research, the same authors introduced Cypider \cite{karbab@Cypider}, a community-based cyber-defense framework that constructs a similarity network of malicious applications and applies community detection algorithms to identify malicious communities without relying on signature-based or learning-based patterns. Cypider demonstrated the viability of community fingerprinting as a means to detect both known malware variants and zero-day threats.

Deep learning has further advanced the state of the art in static malware fingerprinting. MalDozer \cite{karbab2018Maldozer, karbab2017Android} leverages sequences of API method calls to automatically learn malicious and benign patterns, achieving F1-scores between 96\% and 99\% across datasets ranging from 1K to 33K malware samples. Extending this approach with adaptive capabilities, PetaDroid \cite{10.1007/978-3-030-80825-9_16} combines natural language processing and machine learning techniques to achieve resilient Android malware detection and family clustering, maintaining 98--99\% F1-scores while adapting to concept drift and resisting common binary obfuscation techniques. A comprehensive treatment of data-driven Android malware fingerprinting is provided in \cite{karbab2021android}, which consolidates multiple techniques for threat intelligence generation. Beyond Android, SwiftR \cite{karbab2023SwiftR} addresses cross-platform ransomware fingerprinting through a hierarchical neural network architecture that operates on hybrid static and dynamic features, achieving F1-scores up to 98\% on a dataset of 40.3K samples while remaining agnostic to underlying hardware platforms and operating systems. In a related direction, BinEye \cite{alrabaee2019BinEye} demonstrates that deep learning over multiple binary representations—gray-scale images, bytecode sequences, and function-level opcodes—can effectively characterize binary authorship, achieving over 90\% accuracy and revealing authorship relationships in malware families such as Zeus and Citadel.

The most closely related work is MutantX-S \cite{Hu13MutantX}, which also leverages static analysis features to cluster malware samples. MutantX-S uses opcodes extracted from assembly code and employs the feature hashing technique \cite{qinfeng09hashk} to reduce the dimensionality of N-grams from the opcode sequence. The authors adopt a prototype-based algorithm from \cite{Rieck11Automatic} for its efficiency in malware clustering. MutantX-S demonstrated good scalability, clustering 130k samples in 90 minutes with 82\% precision.

\section{Conclusion} 
\label{sec_conclusion}

In this paper, we proposed {\sf TrapNet}, a highly scalable malware clustering framework that combines a set of algorithms and novel techniques to achieve very fast malware clustering. The goal of {\sf TrapNet} is to accelerate the triage process for the massive number of malware samples encountered daily. To do so, {\sf TrapNet} efficiently builds a malware similarity network and then detects highly connected nodes (malware) as suspicious communities. It employs graph community detection over this network to fingerprint the malware communities. We also proposed FloatHash, a novel technique to produce a numerical digest for each malware sample that aims to capture its underlying logic. FloatHash, a short vector of real numbers, is the cornerstone of the malware similarity network in {\sf TrapNet} due to its extremely fast similarity computation. Our evaluation shows that {\sf TrapNet} has excellent scalability on a large dataset of $250,000$ recent wild malware samples. It required only 12 minutes to cluster the dataset, achieving $48\%$ coverage and $82\%$ community purity. {\sf TrapNet} is approximately 15 times faster than MutantX-S \cite{Hu13MutantX} ($130$k in 90 minutes), thereby advancing the state-of-the-art by reducing large-scale malware clustering time from hours to minutes.

\bibliography{references}

\end{document}